\documentclass[%
 reprint,
superscriptaddress,
 amsmath,amssymb,
]{revtex4-2}

\usepackage{graphicx}
\usepackage{dcolumn}
\usepackage{bm}
\usepackage{braket}                             
\usepackage{amsmath}
\usepackage{amsfonts}
\usepackage{bbold}                              
\usepackage{subfigure}                          
\usepackage{hyperref}                           
\usepackage[dvipsnames]{xcolor}                 
\usepackage{multirow}                          
\usepackage{enumitem}                           
\DeclareMathOperator*{\argmin}{arg\,min}  
\usepackage{orcidlink}  

\begin{document}

\title{Quantum optimization with linear Ising penalty functions for customer data science}

\author{Puya~Mirkarimi\,\orcidlink{0000-0001-5835-2592}}
    \email{puya.mirkarimi@durham.ac.uk}
    \affiliation{Department of Physics, Durham University, Durham DH1 3LE, United Kingdom}
\author{Ishaan~Shukla\,\orcidlink{0009-0001-4891-822X}}
    \affiliation{Department of Physics, Durham University, Durham DH1 3LE, United Kingdom}
\author{David~C.~Hoyle\,\orcidlink{0000-0003-3483-5885}}
    \affiliation{dunnhumby, 184 Shepherds Bush Road, London W6 7NL, United Kingdom}
\author{Ross~Williams\,\orcidlink{0000-0001-9347-0922}}
    \affiliation{dunnhumby, 184 Shepherds Bush Road, London W6 7NL, United Kingdom}
\author{Nicholas~Chancellor\,\orcidlink{0000-0002-1293-0761}}
    \email{nicholas.chancellor@gmail.com}
    \affiliation{Department of Physics, Durham University, Durham DH1 3LE, United Kingdom} \affiliation{School of Computing, Newcastle University, 1 Science Square, Newcastle upon Tyne NE4 5TG, United Kingdom}
    
\date{December 16, 2024}

\begin{abstract}
Constrained combinatorial optimization problems, which are ubiquitous in industry, can be solved by quantum algorithms such as quantum annealing (QA) and the quantum approximate optimization algorithm (QAOA). In these quantum algorithms, constraints are typically implemented with quadratic penalty functions. This penalty method can introduce large energy scales and make interaction graphs much more dense. These effects can result in worse performance of quantum optimization, particularly on near-term devices that have sparse hardware graphs and other physical limitations. In this work, we consider linear Ising penalty functions, which are applied with local fields in the Ising model, as an alternative method for implementing constraints that makes more efficient use of physical resources. We study the behaviour of the penalty method in the context of quantum optimization for customer data science problems. Our theoretical analysis and numerical simulations of QA and the QAOA indicate that this penalty method can lead to better performance in quantum optimization than the quadratic method. However, the linear Ising penalty method is not suitable for all problems as it cannot always exactly implement the desired constraint. In cases where the linear method is not successful in implementing all constraints, we propose that schemes involving both quadratic and linear Ising penalties can be effective.
\end{abstract}

\maketitle

\section{\label{sec:introduction}Introduction}

Quantum optimization algorithms, such as quantum annealing (QA)~\cite{Kadowaki1998, Brooke1999, Farhi2001} and the quantum approximate optimization algorithm (QAOA)~\cite{Farhi2014}, are being explored as a potential path to quantum advantage in the near future~\cite{Brady2023, King2024}. There is increasing interest in applying these quantum algorithms to industry problems~\cite{Orus2019, Venturelli2019, Kitai2020, Stollenwerk2020, Fox2021, Yarkoni2021}, including commerce and retail problems~\cite{Nishimura2019, Weinberg2023}. A common feature of optimization problems found in industry is that they are often highly constrained~\cite{Yarkoni2021}. Problems in commerce and retail settings can involve tens to hundreds of constraints~\cite{Czerniachowska2022, Subramanian2010}, which result from both strategic and operational considerations. The large number of constraints influences the approach a researcher might take in tackling an optimization problem. Therefore, if quantum optimization is to find real value in a commercial setting, it must be able to handle and thrive in problems significantly affected by or even dominated by constraints. This motivates the research of quantum optimization on problems that include multiple constraints.

In QA and the QAOA, the standard approach to encoding constraints is to add quadratic penalty functions to a problem's objective function~\cite{Lucas2014, Lodewijks2019}. Quadratic penalty functions can make problems much more dense and introduce large energy scales to energy landscapes. This can negatively impact the performance of a quantum optimizer, especially on near-term devices that have physical limitations~\cite{Preskill2018}. Overcoming these physical limitations requires new strategies that address issues that are not typically a concern in optimization problems. A particular issue that needs to be overcome in quantum annealing is that problems often have to be mapped to quasi-planar hardware graphs, using strategies such as minor embedding~\cite{Choi2008} or parity encoding~\cite{Lechner2015, Rocchetto2016}. Some success has been realised with new encoding strategies, such as domain-wall encoding~\cite{Chancellor2019, Chen2021, Berwald2023}. Domain-wall encoding works well to reduce connectivity when applied to one-hot encodings, but is not efficient for other encodings, while the linear penalty strategy we consider here is.

We consider a penalty method involving only linear Ising terms and study its suitability in quantum optimization for customer data science problems. This penalty method has previously been suggested in other contexts~\cite{Venturelli2019, Ohzeki2020, Willsch2020a}. Unlike the quadratic method, the linear Ising penalty method does not change a problem's connectivity. Furthermore, the energy scales introduced by the linear method are often smaller than for the quadratic method. Because of these reductions in resource usage, theoretical arguments can be made for better performance of quantum optimization with the linear method than the quadratic method. However, while the quadratic penalty method can always exactly implement a desired constraint, there is no such guarantee for the linear Ising penalty method. Thus, the linear Ising penalty method is more applicable to certain problems than others. We find that the customer data science problems we consider are well-suited for the linear method.

In this work, we perform a numerical analysis that pursues two objectives: first, to understand the behaviour of linear Ising penalties, and second, to assess their impact on the performance of quantum optimizer. We study how changing the value of the penalty strength parameter affects the constraint that a linear Ising penalty implements. Strategies for finding a penalty strength that implements the desired constraint are discussed. For problems with multiple constraints, we find that using a combination of quadratic and linear Ising penalties can be beneficial in cases where the linear method is not successful for all constraints. To compare the performance of the penalty methods, we have run simulations of a customer data science problem on QA and the QAOA using both penalty methods. These simulations were limited to fully connected problems with up to 18 variables and did not include the effects of physical limitations that near-term quantum devices typically have. Therefore, many of the theoretical benefits of the linear Ising penalty method do not apply to the setting represented by these simulations. Nevertheless, we observe modest improvements in performance when switching from the quadratic penalty method to the linear Ising penalty method in simulations of both algorithms. These findings are complementary to the experimental results in Ref.~\cite{experimentalpaper}, where more substantial improvements are observed when solving larger customer data science problems on a real quantum device.

This paper is organised as follows. In Sec.~\ref{sec:background}, we outline the relevant prior work, the promotion cannibalization problems we consider, and the linear Ising penalty method. In Sec.~\ref{sec:numerical_methods}, we describe our numerical methods. The results of this study are presented in Sec.~\ref{sec:penalty_functions_behaviour}, where we study the behaviour of linear Ising penalties through a numerical analysis, and Sec.~\ref{sec:simulation_results}, where we compare the performance of quantum optimization using quadratic and linear Ising penalties in simulation. Finally, we summarise our findings in Sec.~\ref{sec:conclusions}.

\section{\label{sec:background}Background}

In this section, we introduce some key concepts and provide a summary of prior works that underpin our study. Sec.~\ref{sec:quantum_optimization} gives a description of QA and the QAOA, which are the two quantum optimization algorithms we consider. Sec.~\ref{sec:promotion_cannibalization_problem} describes the computational problems that we base our numerical analysis on. Sec.~\ref{sec:penalty_methods} reviews the quadratic penalty method for encoding constraints and introduces the linear Ising penalty method.

\subsection{\label{sec:quantum_optimization}Quantum optimization}

QA and the QAOA are quantum algorithms that are particularly suitable for solving combinatorial optimization problems that can be expressed as the quadratic unconstrained binary optimization (QUBO) problem
\begin{align}
    \label{eq:QUBO_objective_function}
    &\text{find} && \argmin_\mathbf{x} f(\mathbf{x}) = \sum_{i=1}^{n} a_i x_i + \sum_{i=1}^{n-1} \sum_{j=i+1}^{n} b_{i,j} x_i x_j.
\end{align}
The real-valued linear term coefficients $\mathbf{a}$ and quadratic term coefficients $\mathbf{b}$ characterise an instance of the problem. An assignment of values to the input vector $\mathbf{x} \in \{0, 1\}^n$ is called a solution, and we are tasked with finding an optimal solution, which is a solution that minimises the objective function $f(\mathbf{x})$. We note that elsewhere in the literature, the QUBO problem is typically expressed in terms of a single upper triangular matrix $Q \in \mathbb{R}^{n \times n}$, where $Q_{i,j} = b_{i,j}~\forall j \neq i$ and $Q_{i,i} = a_i$.

Through the mapping $x_i \mapsto (1 - \sigma_i^z)/2$, the QUBO problem is equivalent to the problem of finding the ground state of the Ising Hamiltonian
\begin{equation}
    H_P = \sum_{i=1}^{n} h_i \sigma_i^z + \sum_{i=1}^{n-1} \sum_{j=i+1}^{n} J_{i,j} \sigma_i^z \sigma_j^z.
    \label{eq:ising_hamiltonian}
\end{equation}
Here, $\sigma_i^z = \mathbb{1}^{\otimes i - 1} \otimes \sigma_z \otimes \mathbb{1}^{\otimes n - i}$ is the Pauli operator $\sigma_z$ acting on qubit $i$ and identities acting on all other qubits. The couplings $\mathbf{J}$ and local fields $\textbf{h}$ can be derived from the QUBO coefficients $\textbf{a}$ and $\textbf{b}$ through the relations
\begin{equation}
    J_{i,j} = \frac{b_{i,j}}{4}
    \label{eq:ising_couplings_qubo_relation}
\end{equation}
and
\begin{equation}
    h_{i} = - \frac{a_i}{2} - \frac{1}{4} \sum_{j=1, j \neq i}^n b_{i,j}.
    \label{eq:ising_field_strengths_qubo_relation}
\end{equation}

QA is a heuristic that resembles simulated annealing~\cite{Kirkpatrick1983}, which is a classical heuristic. It differs to simulated annealing in its use of quantum fluctuations, which are employed through the transverse-field driver Hamiltonian
\begin{equation}
    H_D = -\sum_{i=1}^n \sigma_i^x,
    \label{eq:driver_hamiltonian}
\end{equation}
where $\sigma_i^x$ is defined in terms of the Pauli operator $\sigma_x$ as $\sigma_i^x = \mathbb{1}^{\otimes i - 1} \otimes \sigma_x \otimes \mathbb{1}^{\otimes n - i}$. QA operates by initialising a system in the state $\sum_{j=0}^{2^n-1} \Ket{j} / \sqrt{2^n}$, which is the ground state of $H_D$, and evolving the system according to the Hamiltonian
\begin{equation}
    H(t) = A(t) H_D + B(t) H_P.
    \label{eq:qa_hamiltonian}
\end{equation}
$A(t)$ and $B(t)$ are control functions, which are time-dependent real numbers that satisfy $A(0) \gg B(0)$ at the beginning and $A(t_f) \ll B(t_f)$ at the end of an anneal of duration $t_f$. After this quantum evolution, the system is measured in the computational basis. Our choice of control functions for simulations of QA are $A(t) = 1 - t/t_f$ and $B(t) = t/t_f$, which correspond to a linear schedule. The adiabatic theorem implies that the probability of exciting the system and subsequently measuring an excited state of $H_P$ can be made arbitrarily small by making the anneal time $t_f$ sufficiently large~\cite{Albash2018}. In this work, $t_f$ is not necessarily in the adiabatic regime, so there is some non-negligible probability of sampling a suboptimal solution.

The QAOA is a quantum algorithm that resembles a discretised version of QA. In the QAOA, $p$ layers of gates are applied to an initial state $\Ket{\psi_0}=\sum_{j=0}^{2^n-1} \Ket{j} / \sqrt{2^n}$, which is the same initial state as in QA. In the $k$th layer of gates, the system evolves according to the Ising Hamiltonian $H_P$ for a duration specified by an angle $\gamma_k$ and then by a mixer Hamiltonian $H_M$ for a duration specified by an angle $\beta_k$. The mixer Hamiltonian $H_M$ is a sum of the Pauli operators $\sigma_x$ acting on each qubit, which is the same as the QA driver Hamiltonian $H_D$ in Eq.~\eqref{eq:driver_hamiltonian}. After the application of all $p$ layers of gates, the system is prepared in the state
\begin{equation}
    \Ket{\boldsymbol{\gamma}, \boldsymbol{\beta}} = e^{-i \beta_p H_M} e^{-i \gamma_p H_P} \dots e^{-i \beta_1 H_M} e^{-i \gamma_1 H_P} \Ket{\psi_0},
\end{equation}
which is dependent on $2p$ angles $\boldsymbol{\gamma}$ and $\boldsymbol{\beta}$. The final state $\Ket{\boldsymbol{\gamma}, \boldsymbol{\beta}}$ is measured and produces some solution $\mathbf{x}$ with probability $\left| \braket{\mathbf{x}|\boldsymbol{\gamma}, \boldsymbol{\beta}} \right|^2$. This evolution is a discretised approximation of QA for some choices of the angles $\boldsymbol{\gamma}$ and $\boldsymbol{\beta}$ that can be inferred from the control functions $A(t)$ and $B(t)$. Therefore, the adiabatic theorem also applies to the QAOA in the limit of large $p$, which implies that $\boldsymbol{\gamma}$ and $\boldsymbol{\beta}$ can be chosen such that measuring $\Ket{\boldsymbol{\gamma}, \boldsymbol{\beta}}$ always produces an optimal solution. In practice, $p$ is typically too small to be able to apply the adiabatic theorem. Instead, $\boldsymbol{\gamma}$ and $\boldsymbol{\beta}$ are chosen by a classical optimizer. The optimizer runs the QAOA circuit some number of times and calculates the average objective value $\langle f(\mathbf{x}) \rangle$ of the sampled solutions. It then repeats this process with updated parameters $\boldsymbol{\gamma}$ and $\boldsymbol{\beta}$ in each iteration, using classical techniques to minimise $\langle f(\mathbf{x}) \rangle$.

For a quantum algorithm that produces the state $\Ket{\psi}$ before measurement, the success probability
\begin{equation}
    P_S = \sum_{\mathbf{x}^* \in X^*} \left| \braket{\mathbf{x}^*|\psi} \right|^2
\end{equation}
indicates the probability of measuring an optimal solution, where $X^*$ is the set of optimal solutions. Similarly, the feasible probability
\begin{equation}
    P_F = \sum_{\Tilde{\mathbf{x}} \in \Tilde{X}} \left| \braket{\Tilde{\mathbf{x}}|\psi} \right|^2
\end{equation}
is the probability of measuring a solution from the set $\Tilde{X}$ of solutions that satisfy all constraints.

\subsection{\label{sec:promotion_cannibalization_problem}Promotion cannibalization problem}

In this work, we apply QA and the QAOA to two simplified forms of a problem that arises in customer data science when retailers promote products using price reductions. Promoting a product generates additional revenue from sales of that product; however, it can also have the undesired effect of reducing the sales of other similar products. For example, a promotion of one brand of breakfast cereals may generate new sales for that brand at the expense of other brands' sales. This effect is called cannibalization~\cite{Meredith2001, Aguilar-Palacios2021}, and we will refer to cannibalization arising from product promotions as \textit{promotion cannibalization}. The overall bilateral cannibalization when concurrently promoting two similar products can result in minimal new sales and possibly even a net reduction in revenue. Therefore, retailers often look to minimize the total revenue loss due to clashing concurrent promotions. We can model this by only considering promotion cannibalization that occurs between pairs of products being promoted at the same time, and using a matrix $C$ with matrix elements $C_{i,j}$ that represent the average amount of loss of revenue from sales of product $i$ due to a promotion of product $j$ when both products are promoted concurrently. In this study, we make the assumption that $C$ is nonnegative, that is, $C_{i,j} \geq 0~\forall i, j$.

The primary example problem that we base this work on is to choose a selection of $A$ products to promote out of $n_p$ possible choices that minimises the total amount of promotion cannibalization between pairs of promoted products. This problem can be expressed as the constrained binary quadratic programming problem~\cite{nocedal1999numerical, VanThoai2013}
\begin{align}
    \label{eq:single_quarter_promotion_cannibalization_qubo_objective_function}
    &\text{find} && \argmin_\mathbf{x} f(\mathbf{x}) = \sum_{j=1}^{n_p} \sum_{i=1}^{n_p} C_{i,j} x_i x_j \\
    \label{eq:constraint_C1_single_quarter}
    &\text{s.t.} && \sum_{i=1}^{n_p} x_i = A,
\end{align}
where the binary variable $x_i = 1$ if product $i$ is promoted and $x_i = 0$ if not. Eq.~\eqref{eq:constraint_C1_single_quarter} is an expression of the constraint that $A$ products are promoted in total. Note that when expressed in the form of Eq.~\eqref{eq:QUBO_objective_function}, the objective function only has quadratic terms with coefficients $b_{i,j}$ equal to $C_{i,j} + C_{j,i}$. Hence, we can assume that $C$ is symmetric in the context of this problem without loss of generality. Conversions from QUBO to Ising form are performed using Eq.~\eqref{eq:ising_couplings_qubo_relation} and Eq.~\eqref{eq:ising_field_strengths_qubo_relation}. For example, a $C$ matrix with a single pair of nonzero elements $C_{i,j} = C_{j,i} = 1$ would produce a QUBO objective function containing a single quadratic term with coefficient $b_{i,j} = C_{i,j} + C_{j,i} = 2$, which would correspond to $J_{i,j} = \frac{1}{2}$ and $h_i = h_j = -\frac{1}{2}$ in Ising form.

Retailers change which products are promoted periodically, such as once every fiscal quarter. The above problem can be viewed as finding the promotion plan for a single quarter of the year. Another problem that we consider is to find an optimal promotion plan for two consecutive quarters, subject to the additional set of constraints that each product is promoted in one of the quarters at most. The optimal promotion plan should minimise the total amount of promotion cannibalization between pairs of promotions in the same quarter, and each quarter should have $A$ promotions. This can be expressed as
\begin{align}
    \label{eq:two_quarter_promotion_cannibalization_qubo_objective_function}
    &\text{find} && \argmin_\mathbf{x} f(\mathbf{x}) = \sum_{q=1}^2 \sum_{j=1}^{n_p} \sum_{i=1}^{n_p} \lambda_q C_{i,j} x_{i,q} x_{j,q}, \\
    \label{eq:constraint_C1}
    &\text{s.t.} && \sum_{i=1}^{n_p} x_{i,q} = A~~\forall q & \\
    \label{eq:constraint_C3}
    &\text{and} && x_{i,1} + x_{i,2} \leq 1~~\forall i.
\end{align}
Here, the binary variable $x_{i,q}$ represents a promotion of product $i$ in quarter $q$ and $\lambda_q$ is a seasonal scale factor that is derived from the expected total revenue of each quarter. In total, this problem has $n_p + 2$ constraints to satisfy, which allows us to study the behaviour of interacting constraints. For clarity, we refer to Eqs.~\eqref{eq:single_quarter_promotion_cannibalization_qubo_objective_function}--\eqref{eq:constraint_C1_single_quarter} as the \textit{single-quarter} promotion cannibalization problem and Eqs.~\eqref{eq:two_quarter_promotion_cannibalization_qubo_objective_function}--\eqref{eq:constraint_C3} as the \textit{two-quarter} problem.

\subsection{\label{sec:penalty_methods}Penalty methods for encoding constraints}

The promotion cannibalization problems, as expressed in Eqs.~\eqref{eq:single_quarter_promotion_cannibalization_qubo_objective_function}--\eqref{eq:constraint_C1_single_quarter} and Eqs.~\eqref{eq:two_quarter_promotion_cannibalization_qubo_objective_function}--\eqref{eq:constraint_C3}, are not QUBO problems because they include constraints. Eq.~\eqref{eq:constraint_C1_single_quarter} and Eq.~\eqref{eq:constraint_C1} are examples of linear equality constraints, which take the general form
\begin{equation}
    \sum_{i=1}^n \mu_i x_i = c,
    \label{eq:linear_equality_constraint}
\end{equation}
for some coefficients $\boldsymbol{\mu} \in \mathbb{R}^n$ and constraint value $c \in \mathbb{R}$. The most common method to encode constraints in quantum optimization is to incorporate them into the objective function through the addition of penalty functions $P(\textbf{x})$ that raise the objective value of solutions that are infeasible (i.e.\ don't satisfy every constraint) enough that the solution with the lowest objective value is feasible.

Quantum optimizers and classical quadratic programming solvers commonly use the quadratic penalty function
\begin{equation}
    P(\mathbf{x}) = \alpha_2 \left(\sum_{i=1}^n \mu_i x_i - c \right)^2
    \label{eq:quadratic_penalty_function}
\end{equation}
to encode equality constraints of the form given in Eq.~\eqref{eq:linear_equality_constraint}. Provided that the penalty strength $\alpha_2$ is large enough, the optimal solution of an objective function will satisfy Eq.~\eqref{eq:linear_equality_constraint} after adding $P(\textbf{x})$ to it. Aside from the ability to scale the function by changing the penalty strength $\alpha_2$, this penalty function has two desirable properties that are satisfied for all $\alpha_2 > 0$:
\begin{enumerate}
    \item $P(\textbf{x}) = 0$ if $\textbf{x}$ is feasible.
    \label{item:penalty_function_property_1}
    \item $P(\textbf{x}) > 0$ if $\textbf{x}$ is infeasible.
    \label{item:penalty_function_property_2}
\end{enumerate}

The quadratic penalty method suffers from severe drawbacks. Expanding out the brackets in Eq.~\eqref{eq:quadratic_penalty_function} for a quadratic penalty, we find that there are quadratic terms with nonzero coefficients for all pairs of variables in the constraint. This corresponds to all-to-all couplings of the associated qubits in the Ising Hamiltonian $H_P$. Many quantum devices that are currently available do not support all-to-all couplings between the physical qubits, meaning that most Ising Hamiltonians of interest cannot be directly mapped to the hardware. To resolve this issue, various methods can be used. In the gate-based setting of QAOA, the quantum states of two qubits can be swapped with a SWAP gate, allowing for any two logical qubit states to be routed through the hardware so they can be physically coupled~\cite{Kivlichan2018, OGorman2019, Hagge2020, Hashim2022}. In QA, a common approach is to map each logical qubit to a chain of ferromagnetically coupled physical qubits in a process called minor embedding~\cite{Choi2008, Yarkoni2021} such that the chains of qubits support the necessary couplings. In both cases, a significant physical resource cost is incurred when new couplings are introduced to $H_P$, either in the form of a larger circuit depth in the case of the QAOA or a larger qubit count in the case of QA.

As well as the cost of requiring all-to-all couplings, another drawback of the quadratic penalty method is that it will often reduce the effective dynamic range of qubit interactions. The quadratic penalty for the constraint in Eq.~\eqref{eq:constraint_C1_single_quarter} is
\begin{equation}
    P(\mathbf{x}) = \alpha_2 \left(\sum_{i=1}^{n_p} x_{i, q} - A \right)^2.
    \label{eq:C1_quadratic_penalty_function}
\end{equation}
Expanding out the brackets, we get
\begin{equation}
    P(\mathbf{x}) = \alpha_2 \left( \sum_{i=1}^{n_p}(1-2A)x_i + \sum_{i=1}^{n_p-1} \sum_{j=i+1}^{n_p} 2 x_i x_j + A^2 \right).
    \label{eq:C1_quadratic_penalty_function_expanded}
\end{equation}
Mapping this to an Ising Hamiltonion with $x_i \mapsto (1 - \sigma_i^z)/2$ gives
\begin{multline}
    P = \alpha_2 \left( \sum_{i=1}^{n_p-1} \sum_{j=i+1}^{n_p} \frac{\sigma_i^z \sigma_j^z}{2} \right. \\
    \left. + \sum_{i=1}^{n_p} \left( \frac{n_p}{2} - A \right) \sigma_i^z + \frac{n_p(n_p + 1)}{4} - n_p A + A^2 \right) .
\end{multline}
Now, we can read off the couplings and local fields as $J_{i,j} = \alpha_2 / 2$ and $h_i = \alpha_2 (n_p/2 - A)$. As we would expect, the magnitudes of $\mathbf{J}$ and $\mathbf{h}$ increase with the magnitude of $\alpha_2$. As well as this, the magnitude of $\mathbf{h}$ is proportional to the absolute difference $|n_p/2 - A|$. In other words, this penalty introduces strong local fields if the desired constraint value (number of promotions $A$) is far from half of the number of variables in the constraint (number of products $n_p$).

In QA, large-magnitude couplings or local fields are undesirable because there are physical limitations on the range of $\mathbf{J}$ or $\mathbf{h}$ values that can be implemented. The Ising Hamiltonian $\Tilde{H}_P$ implemented by a quantum annealer is normalised by a factor $\mathcal{N}$ using
\begin{equation}
    \Tilde{H}_P = \frac{1}{\mathcal{N}} H_P,
    \label{eq:normalised_ising_hamiltonian}
\end{equation}
where $\mathcal{N}$ is typically chosen to be the minimum value such that all physical constraints on $\mathbf{J}$ and $\mathbf{h}$ are satisfied. A penalty that increases the maximum magnitude values in $\mathbf{J}$ or $\mathbf{h}$ will often result in a larger $\mathcal{N}$, which reduces the effective dynamic range of qubit interactions for the unconstrained part of the problem. Large energies associated with penalties would also be detrimental in the QAOA because the application of $H_P$ would begin rotating phases by angles larger than $2 \pi$ unless the rotation angles $\gamma$ are reduced or $H_P$ is normalised. This restricts the effective range of rotation angles for the interactions in $H_P$ that are not associated with the penalties.

Various alternatives to the quadratic penalty method have also been explored in the context of quantum optimization. In QA, one approach is to start in a superposition of feasible states and use an alternative driver Hamiltonian $H_D$ that only drives transitions between feasible states~\cite{Hen2016, Hen2016a}. This can also be applied to a generalisation of the QAOA by replacing $H_M$ with unitaries that implement the constraints~\cite{Hadfield2019}. While this method is appealing as it naturally limits the quantum evolution to the subspace of feasible states, it requires multi-qubit interactions that are more difficult to physically implement than the transverse-field Hamiltonian. This method can be combined with a parity-based encoding scheme~\cite{Lechner2015, Drieb-Schon2023}, which presents an alternative to minor embedding in QA for compatible hardware. Another approach is to define small Ising problems called \textit{gadgets} that have certain properties that allow them to be combined to encode the original constrained problem~\cite{Vyskocil2019, Vyskocil2019a, Djidjev2020}. These gadgets use ancillary qubits, which is the same type of resource cost as for the quadratic penalty method when considering the extra qubits introduced in minor embedding. When tailored to a device's hardware, the gadget-based approach can be more efficient than the quadratic penalty method in terms of dynamic range and the number of physical qubits required.

In classical computing, penalty functions that are linear in $\textbf{x}$ are also used~\cite{Fletcher1983}. To satisfy property~\ref{item:penalty_function_property_2}, non-Ising operations are required. For example, the constraint in Eq.~\eqref{eq:linear_equality_constraint} can be implemented with the linear penalty $\alpha_1 \left| \sum_{i=1}^n \mu_i x_i - c  \right|$.
Computing using non-Ising operations such as $|\cdot|$ is challenging in quantum optimization. de la Grand'rive and Hullo~\cite{DelaGrandrive2019} demonstrated that the QAOA can be extended to implement non-Ising linear penalties for inequality constraints by computing $\max(\cdot)$, which is a non-Ising operation. However, this approach comes at the cost of requiring ancillary qubits and a more complex circuit.

In this paper, we consider the use of linear penalty functions with non-Ising operations removed. This method would be able to avoid many of the drawbacks of the quadratic penalty method as it does not introduce any new couplings to the problem and can be more efficient with the hardware's dynamic range~\cite{Ohzeki2020, Yu2021}. The removal of non-Ising operations requires giving up on property~\ref{item:penalty_function_property_2} of a conventional penalty function, which means this penalty method does not produce the desired ground state in $H_P$ for all cases. The upside is that this penalty method can be implemented on current quantum hardware with a smaller physical overhead than other methods and may potentially improve the performance of algorithms. This type of linear penalty method has been previously suggested for QA in the contexts of portfolio optimization~\cite{Venturelli2019} and quantum machine learning~\cite{Willsch2020a}, but neither of these two studies showed any results relating to this approach beyond making the observation that it can be used in these problems.

Ohzeki~\cite{Ohzeki2020} recently developed a method for implementing constraints with linear terms by taking the partition function of a QUBO objective function and applying a Hubbard-Stratonovich transformation~\cite{stratonovich1957method, Hubbard1959}. While the mathematical motivation behind Ohzeki's work is different to what we present here, the resulting algorithm is effectively the same as applying linear Ising penalties. It has been observed that the method described by Ohzeki does not have the theoretical guarantee of being able to exactly implement all hard constraints~\cite{Kuramata2021}. This work examines objective functions that cannot be exactly constrained by this method and considers their implications, which the prior work does not do. We discuss how the structure of a problem affects the likelihood of this occuring and identify that objective functions with non-negative quadratic term coefficients are particularly suitable for the linear Ising penalty method. One proposal for making use of infeasible solutions is to perform a post-processing step to produce feasible solutions from infeasible solutions using the fewest number of bit flips~\cite{Kuramata2021}. In this work, we propose another strategy of selectively applying linear Ising penalties to specific constraints when it is not possible to apply them to all of a problem's constraints. Whereas Ohzeki and subsequent researchers performed their analyses on D-Wave quantum annealing devices, this work presents results of numerical simulations in both gate-based and annealing settings. Performing simulations allows us to study the behaviour of closed-system dynamics and to study how the dynamics are influenced by the choice of penalty strength parameters.

The linear Ising penalty function for the equality constraint in Eq.~\eqref{eq:constraint_C1_single_quarter} is
\begin{equation}
    P(\mathbf{x}) = \alpha_1 \left( \sum_{i=1}^{n_p} x_{i,q} - A \right),
    \label{eq:C1_linear_penalty_function}
\end{equation}
where the penalty strength $\alpha_1$ can be positive or negative. In the Ising formulation, this corresponds to local fields equal to $-\alpha_1/2$ up to an unimportant constant offset. In the rest of this paper, when using the term \textit{linear penalty}, we refer to linear Ising penalties of the form in Eq.~\eqref{eq:C1_linear_penalty_function} rather than linear penalties with non-Ising operations, which are used in classical computing.

Since we assume that $C$ is non-negative, the QUBO objective function of the single-quarter promotion cannibalization problem only has quadratic terms with non-negative coefficients. To see how this affects the problem's structure, we group all possible solutions by their Hamming weight $w$, which represents the number of variables equal to $1$ in a solution. The minimum objective value of the solutions for each group is monotonically increasing with $w$. In other words, for every solution with some Hamming weight $w \geq 1$, there exists some solution with Hamming weight $w-1$ that has the same or a lower objective value. To see why this is true, suppose there exists a solution $\mathbf{x}_a$ with objective value $f(\mathbf{x}_a)$ and Hamming weight $w$, where $f(\mathbf{x}_a)$ is strictly less than the objective values of all solutions with Hamming weight $w-1$. We can flip one of the variables of $\mathbf{x}_a$ from 1 to 0 to get another solution $\mathbf{x}_b$ with a Hamming weight $w-1$. Since the QUBO objective function does not have any terms with negative coefficients, flipping a variable from 1 to 0 can never increase the objective value. Hence, $f(\mathbf{x}_b) \leq f(\mathbf{x}_a)$, which contradicts the proposition that $f(\mathbf{x}_a)$ is strictly less than the objective value of all solutions with Hamming weight $w-1$. Therefore, solutions with a Hamming weight $w$ can never have an objective value that is strictly less than the objective value of all solutions with Hamming weight $w-1$. The same is true for the two-quarter promotion cannibalization problem when considering the Hamming weight of the variables for one quarter while the other quarter's variables have fixed values.

\begin{figure*}
    \includegraphics[width=1.4\columnwidth]{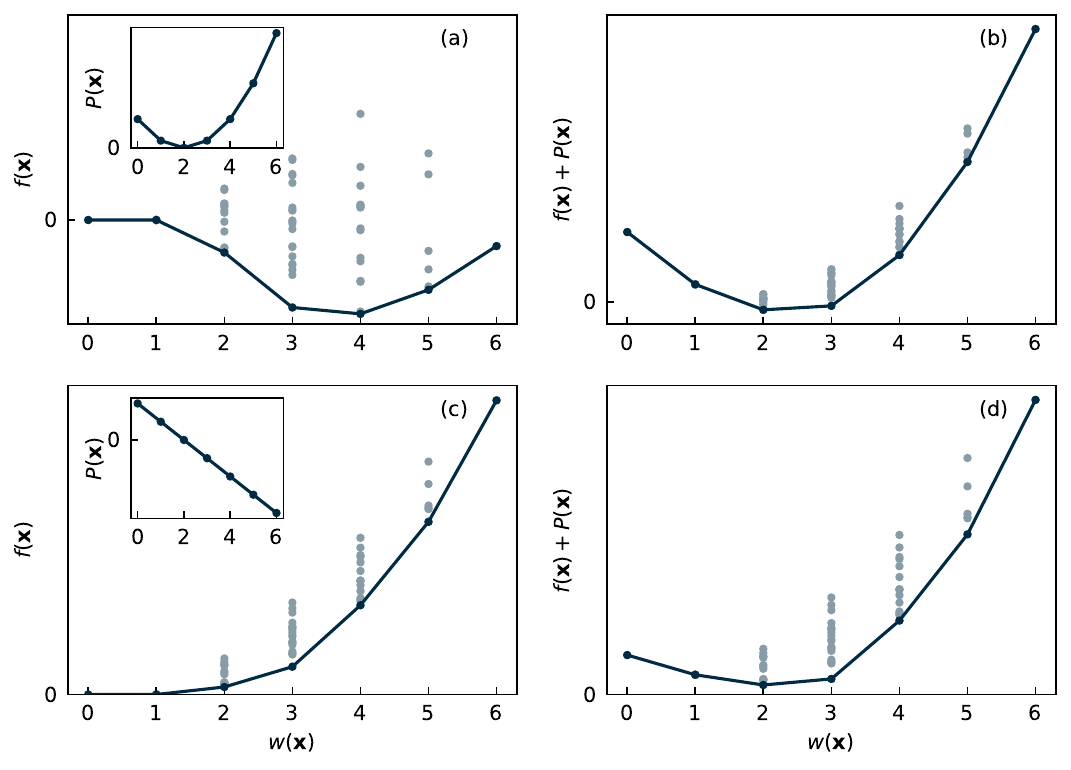}
    \caption{Objective values $f(\mathbf{x})$ plotted against the Hamming weight $w(\mathbf{x})$ of each possible solution $\mathbf{x}$ for (a) an example QUBO instance with both positive and negative quadratic term coefficients and (c) an instance of the single-quarter promotion cannibalization problem. For each Hamming weight, the minimum value of $f(\mathbf{x})$ is plotted in a darker colour, and these points are connected to guide the eye. We apply two different penalty functions $P(\mathbf{x})$ for the equality constraint $\sum_{i=1}^6 x_i = 2$ to these problem instances, which are shown in the insets of (a) and (c). The penalty function is quadratic in (a) and linear in (c). Note that the gradient of the line in the inset of (c) is equal to $\alpha_1$. (b) The constrained objective function after applying a quadratic penalty to the QUBO instance plotted in (a). (d) The constrained objective function after applying a linear penalty to the promotion cannibalization problem instance plotted in (c).}
    \label{fig:penalty_methods_diagrams}
\end{figure*}

The monotonic relationship between Hamming weight and minimum objective value makes this problem particularly amenable to the linear Ising penalty method. To demonstrate this, we plot $f(\mathbf{x})$ against the Hamming weight of $\mathbf{x}$ for a promotion cannibalization problem instance with six products in Fig~\ref{fig:penalty_methods_diagrams}(c) and for some other six-variable QUBO instance with both positive and negative quadratic term coefficients in Fig~\ref{fig:penalty_methods_diagrams}(a). If we suppose we want to implement the constraint $\sum_{i=1}^6 x_i = 2$ for both of these problems, the penalised objective functions must have minimum values at a solution Hamming weight of two. For the promotion cannibalization problem, the minimum value of $f(\mathbf{x})$ is monotonically increasing with the Hamming weight of $\mathbf{x}$, so only solutions with a Hamming weight of less than two need to be penalised to implement the desired constraint. A linear penalty (shown in the inset of Fig~\ref{fig:penalty_methods_diagrams}(c)) is able to do this, but it also has the undesired effect of lowering the objective value of solutions with Hamming weights greater than two. However, this is often compensated by the large existing object values of these higher Hamming weight solutions due to the monotonic structure we described. Therefore, the linear penalty method is more likely to be able to produce the correct optimal solution than for a random QUBO problem without the same structure. Indeed, we find that for this instance of the promotion cannibalization problem, the linear penalty method is successful in producing a constrained objective function, which is shown in Fig.~\ref{fig:penalty_methods_diagrams}(d). For the other QUBO problem, there does not exist any value of $\alpha_1$ that produces an optimal solution with a Hamming weight of two, so the quadratic penalty method is used instead (Fig.~\ref{fig:penalty_methods_diagrams}(a) and Fig.~\ref{fig:penalty_methods_diagrams}(b)). We note that the monotonic structure of the promotion cannibalization problem is not sufficient to guarantee that the linear penalty method is successful every time. For that, the gradient of the line in Fig~\ref{fig:penalty_methods_diagrams}(c) also needs to be monotonically increasing, which isn't the case for all problem instances.

The constraints given by Eq.~\eqref{eq:constraint_C3} are examples of linear inequality constraints. We encode these with quadratic penalty functions of the form
\begin{equation}
    P(\mathbf{x}) = \alpha_2 x_{i,1} x_{i,2}
    \label{eq:C2_quadratic_penalty_function}
\end{equation}
throughout our analysis of the two-quarter promotion cannibalization problem.

\section{\label{sec:numerical_methods}Numerical methods}

The numerical analysis in this work was performed using the Python programming language~\cite{VanRossum1995}. We used the libraries NumPy~\cite{harris2020array} and SciPy~\cite{2020SciPy-NMeth} for computationally intensive calculations. All linear fits were obtained using the weighted least-squares method implementation in \texttt{scipy.optimize.curve\_fit}. PyQUBO~\cite{Zaman2022} was used to formulate QUBO and Ising problems. Plots were produced with Matplotlib~\cite{hunter2007matplotlib}.

Problem instances were exactly solved using Gurobi Optimizer~\cite{gurobi} through the GurobiPy Python interface. We note that Gurobi operates at an adjustable numerical precision, which can lead to minor differences in results depending on which software version and solver parameters are used. Throughout this work, we used Gurobi version 10.0.2 with a single thread and the default values of all other solver parameters.

The $C$ matrices used in our numerical analysis of the single-quarter promotion cannibalization problem were generated by selecting symmetric off-diagonal matrix elements $C_{i,j} = C_{j,i}$ uniformly at random from the interval $[0.1, 1.0)$. All matrix elements on the main diagonal were set to $0$. With this method, 
we generated 10,000 $C$ matrices corresponding to different problem instances for each number of products between 6 and 18. The $C$ matrices of the instance with ID \texttt{8\_0} and the instance with ID \texttt{8\_19} are also used for our analysis of the two-quarter promotion cannibalization problem.

All simulations in this work were run on the Hamilton high performance computing cluster at Durham University. The SciPy function \texttt{expm\_multiply} was used to simulate QA by discretising the time evolution in the same manner as described in Ref.~\cite{Festenstein}. For the QAOA, the \texttt{qasm\_simulator} backend in the Qiskit SDK~\cite{Qiskit} was used to simulate quantum circuits without noise. Each of the QAOA angles $\boldsymbol{\gamma}$ and $\boldsymbol{\beta}$ were initialised randomly within the interval $[0, 1)$ before being optimized by the implementation of the COBYLA algorithm~\cite{Powell1994} in \texttt{scipy.optimize.minimize} using a maximum of 100 minimisation iterations. While optimizing $\boldsymbol{\gamma}$ and $\boldsymbol{\beta}$, the quantum circuit was run 1,000 times in order to estimate the average objective value. After optimizing $\boldsymbol{\gamma}$ and $\boldsymbol{\beta}$, 1,000,000 runs of the quantum circuit were performed, from which the probabilities $P_S$ and $P_F$ were inferred. In order to reduce the effect of getting stuck in local minima when optimizing $\boldsymbol{\gamma}$ and $\boldsymbol{\beta}$, the entire algorithm was repeated 80 times before selecting the attempt that produced the highest success probability.

\section{\label{sec:penalty_functions_behaviour}Behaviour of penalty functions}

In this section, we study the behaviour of linear and quadratic penalty functions as their penalty strength parameters are changed. We first look at the behaviour of a single constraint using the single-quarter promotion cannibalization problems as an example. Then, we use the two-quarter promotion cannibalization problem to consider the effects of having multiple constraints.

\subsection{Single constraint}

\begin{figure}
    \centering
    \includegraphics[width=\columnwidth]{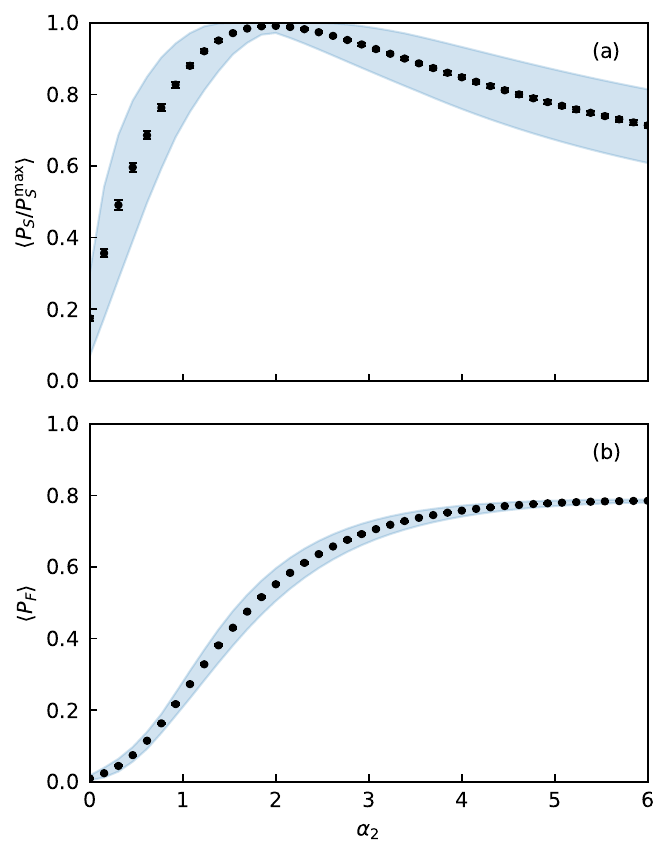}
    \caption{(a) Probability $P_S$ after an anneal time $t_f=10$ of measuring the optimal state normalised by an estimate of the maximum success probability $P_S^\mathrm{max}$ for closed-system QA as a function of the penalty parameter $\alpha_2$. Each point is an average over 100 instances of the promotion cannibalization problem with $n_p=12$ products, with error bars representing the standard error in the mean. The blue shaded region contains the 5th to 95th percentile values. Note that $P_S^\mathrm{max}$ is calculated for each instance separately and is estimated by taking the maximum $P_S$ over the plotted values of $\alpha_2$. (b) Same as (a) with the y-axis instead showing the probability $P_F$ of measuring a feasible state.}
    \label{fig:qa_quadratic_penalty_strength}
\end{figure}

While both the quadratic and linear penalty methods introduce a single penalty strength parameter per constraint, these two types of penalties respond differently to changes in their penalty strengths. In this subsection, we study these differences in the context of the single-quarter promotion cannibalization problem, which has a single constraint (Eq.~\eqref{eq:constraint_C1_single_quarter}). For the quadratic method, if the penalty strength $\alpha_2$ is too small, it is not successful in producing a feasible ground state in $H_P$. On the other hand, if $\alpha_2$ is much larger than necessary, $H_P$ will have a feasible ground state, but the performance of a quantum optimizer in finding the ground state will be hindered. Fig.~\ref{fig:qa_quadratic_penalty_strength} shows an example of this behaviour for simulations of closed-system QA. The feasible probability $P_F$ increases with $\alpha_2$ and stays elevated at large values, whereas the success probability $P_S$ rises, peaks, and then falls. For a given problem instance, there is a value of $\alpha_2$ that will produce the maximum probability $P_S^\mathrm{max}$ of measuring an optimal solution. In practice, there is typically a broad range of values for $\alpha_2$ that will produce good performance, which can be seen in Fig.~\ref{fig:qa_quadratic_penalty_strength}(a).

\begin{figure}
    \centering
    \includegraphics[width=\columnwidth]{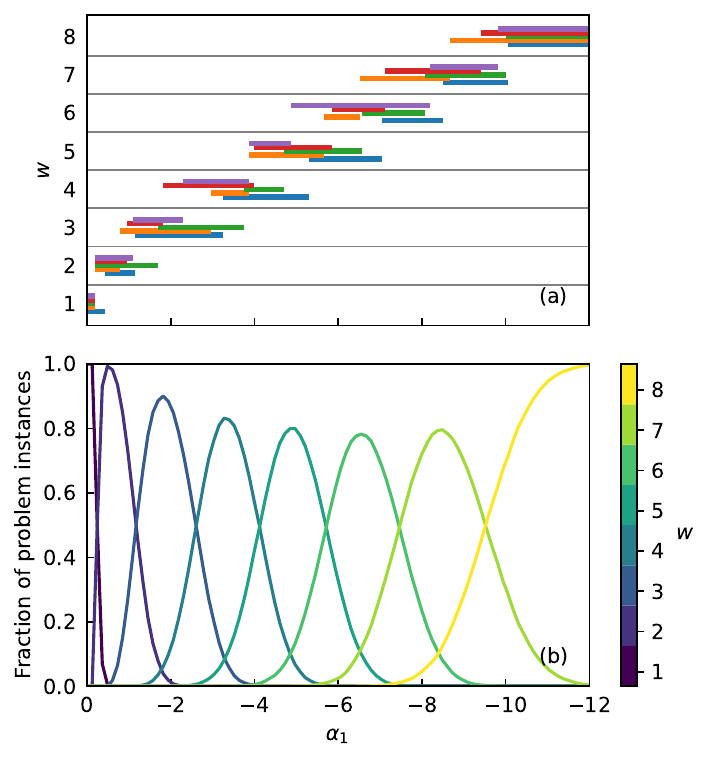}
    \caption{(a) The Hamming weight $w$ of the ground state of $H_P$ plotted against the linear penalty strength $\alpha_1$ for five random instances of the promotion cannibalization problem with 8 products. Each instance corresponds to a different coloured bar. (b) The fraction of 10,000 single-quarter problem instances that have a ground state Hamming weight of $w$ plotted against $\alpha_1$. Note that $\alpha_1$ is decreasing along the x-axis.}
    \label{fig:linear_penalty_hamming_weight}
\end{figure}

For the linear penalty method, the parameter $\alpha_1$ directly relates to the value of the constraint that is implemented, i.e.\ the value of $A$ in Eq.~\eqref{eq:constraint_C1_single_quarter}. As shown in Fig.~\ref{fig:linear_penalty_hamming_weight}(a), the Hamming weight $w$ of the ground state of $H_P$ increases as $\alpha_1$ is decreased. All instances of the Ising problem exhibit this monotonic relation between $\alpha_1$ and $w$ because negative local fields lower the energy of states in proportion to their Hamming weights. For a desired value of $A$, the penalty strength $\alpha_1$ should be tuned such that $w=A$. There is a lower and upper limit on the range of values of $\alpha_1$ that accomplish this, and this range is different for each problem instance. In comparison, there is only a lower limit for $\alpha_2$ that produces a feasible ground state when using the quadratic penalty method. Therefore, the range of values of $\alpha_1$ that produce a significant probability of finding an optimal solution is often smaller than the equivalent range of values of $\alpha_2$. Furthermore, for some problem instances, there does not exist any value of $\alpha_1$ that produces the desired ground state in $H_P$ with the linear penalty method, which is not the case for the quadratic method.

When choosing $\alpha_1$, while there may exist a value that implements the desired constraint in the ground state of $H_P$ for most instances of a given problem, there typically won't be a single value that works for all of these instances, as can be seen in Fig.~\ref{fig:linear_penalty_hamming_weight} for the single-quarter promotion cannibalization problem. However, the monotonic relationship between $w$ and $\alpha_1$ means that we can iteratively search for a good value of the parameter by decreasing (increasing) $\alpha_1$ when sampled solutions have too few (many) ones. This has previously been suggested in Refs.~\cite{Venturelli2019, Ohzeki2020}. For the single-quarter promotion cannibalization problem, an effective strategy is to start with a guess of $\alpha_1=-1$, run the optimizer, and double $\alpha_1$ until the returned solution has the correct Hamming weight $w=A$ or one that is too large. In the case where $w$ is larger than $A$, a binary search can be performed to find a value of $\alpha_1$ that produces the desired value of $w$. Assuming the optimizer always returns the optimal solution, this method can find $\alpha_1$ in a number of calls to the optimizer that scales logarithmically with the value of $\alpha_1$ that correctly implements the constraint and is closest to the initial guess of $-1$. The monotonic relationship between $\alpha_1$ and $w$ is the condition that makes this true. There is a possibility of the binary search terminating before it finds a correct value of $\alpha_1$ if the interval of correct values of $\alpha_1$ is smaller than the precision of the search. The search can be made exponentially more precise by increasing the number of binary search iterations. See Ref.~\cite{experimentalpaper} for an example algorithm that iteratively searches for good values of $\alpha_1$ for problems with one or more linear penalties. The iterative nature of the application of linear penalties makes this method technically a hybrid quantum-classical technique~\cite{Callison2022}.  

\begin{figure}
    \centering
    \includegraphics[width=\columnwidth]{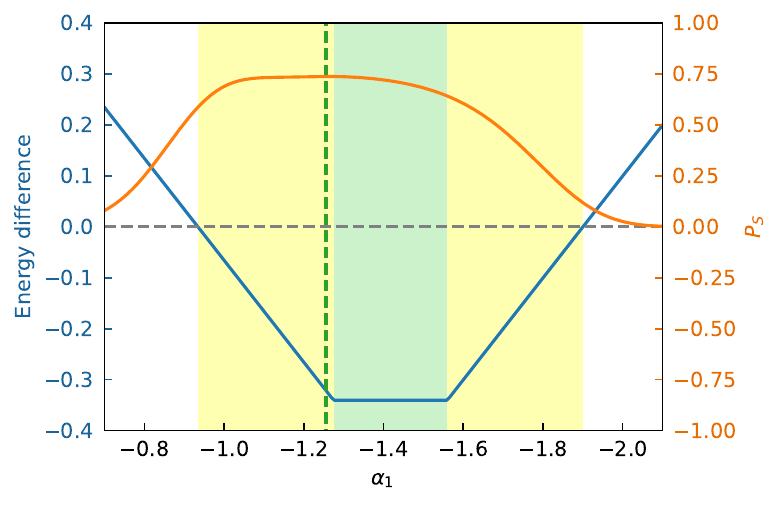}
    \caption{
    Behaviour of the linear penalty method for a random single-quarter problem instance with 10 products and a constraint of $A=3$ promotions. In orange, we plot the success probability $P_S$ of finding the optimal feasible solution $\mathbf{x}^*$ after an anneal time $t_f=200$ for closed-system QA. Note that the penalty strength $\alpha_1$ is decreasing along the x-axis. In blue, we plot the energy difference between the desired state $\ket{\mathbf{x}^*}$ and the minimum energy eigenstate of $H_P$ that isn't $\ket{\mathbf{x}^*}$, where a negative energy difference (highlighted by the yellow and green shaded regions) indicates that $\ket{\mathbf{x}^*}$ is the non-degenerate ground state of $H_P$. The green shaded region indicates where $\ket{\mathbf{x}^*}$ is the ground state and the energy separation to the first excited state is maximised. The $\alpha_1$ value that maximises $P_S$ is indicated by the green dashed line.}
    \label{fig:linear_penalty_energy_and_success_prob}
\end{figure}

Not only does $\alpha_1$ need to be within a finite interval to implement the desired constraint, but the specific choice of $\alpha_1$ within that interval impacts the performance of the quantum algorithm. Fig.~\ref{fig:linear_penalty_energy_and_success_prob} demonstrates this for one particular promotion cannibalization problem instance. While all choices of $\alpha_1$ in the shaded regions of the plot produce the correct ground state in $H_P$, some choices result in significantly better success probabilities than others. Since in practice the optimal choice of $\alpha_1$ is not known in advance, our analysis of the performance of the linear penalty method in Sec.~\ref{sec:simulation_results} uses values that have been selected uniformly at random from the interval that produces the correct ground state in $H_P$. Fig.~\ref{fig:linear_penalty_energy_and_success_prob} also shows that it is possible for QA to have a significant probability of measuring the optimal feasible solution even when the choice of $\alpha_1$ does not make the ground state feasible (i.e.\ outside the shaded regions). This probability will tend to zero as the anneal time is increased towards the adiabatic limit.

\subsection{Multiple constraints}

In applied optimization, it is common for problems to have many constraints. Therefore, it is important to consider how linear penalty functions interact with each other. We use the two-quarter promotion cannibalization problem to study this. The seasonal scale factors in Eq.~\eqref{eq:two_quarter_promotion_cannibalization_qubo_objective_function} are set to $\boldsymbol{\lambda}=(1.5, 1.0)^\mathrm{T}$ in the problem instances we consider. For the ground state of $H_P$, we denote the Hamming weights of the variables associated with the first and second fiscal quarters as $w_1$ and $w_2$ respectively, which correspond to the number of promotions in each quarter. The constraints in Eq.~\eqref{eq:constraint_C1} require that $w_1 = w_2 = A$.

\begin{figure}
    \centering
    \includegraphics[width=\columnwidth]{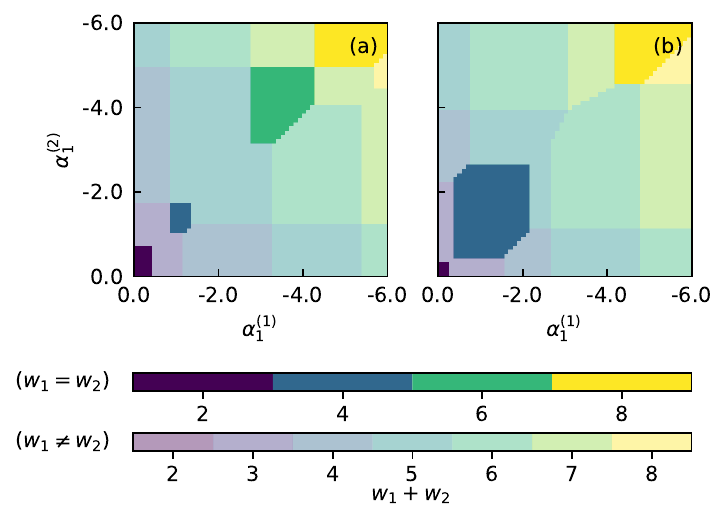}
    \caption{For the two-quarter promotion cannibalization problem with $n_p=8$ products, linear penalties are applied to the first and second quarters with penalty strengths $\alpha_1^{(1)}$ and $\alpha_1^{(2)}$ respectively. This produces a ground state of $H_P$ that has a first-quarter Hamming weight $w_1$ and second-quarter Hamming weight $w_2$. Heat maps show the sum $w_1 + w_2$ as a function of the penalty strengths for the $C$ matrices of (a) the instance with ID \texttt{8\_0} and (b) the instance with ID \texttt{8\_19}. Two different colour bars are used depending on whether $w_1$ and $w_2$ are equal or not. More saturated colours are used where $w_1 = w_2$ and less saturated colours are used where $w_1 \neq w_2$. Note that $\alpha_1^{(1)}$ and $\alpha_1^{(2)}$ are decreasing along the axes of the plots.}
    \label{fig:linear_penalties_colour_map}
\end{figure}

We first consider the case where linear penalty functions are used for both constraints in Eq.~\eqref{eq:constraint_C1}, with the penalty strengths for the constraints on the first and second quarters denoted as $\alpha_1^{(1)}$ and $\alpha_1^{(2)}$ respectively. Fig.~\ref{fig:linear_penalties_colour_map} shows how $w_1$ and $w_2$ change with $\alpha_1^{(1)}$ and $\alpha_1^{(2)}$ for two problem instances. In Fig.~\ref{fig:linear_penalties_colour_map}(a), there are regions where $w_1 = w_2 = A$ is satisfied for each possible value of $A$ that also satisfies Eq.~\eqref{eq:constraint_C3}. However, finding these regions is not as simple as using the same strategy that can be used for the single-quarter problem to search for $\alpha_1^{(1)}$ and $\alpha_1^{(2)}$ independently. This is because changes in $\alpha_1^{(1)}$ impact both $w_1$ and $w_2$, and similarly for $\alpha_1^{(2)}$. Therefore, $\alpha_1^{(1)}$ and $\alpha_1^{(2)}$ cannot be found independently of each other. The monotonic relationship between $w_i$ and $\alpha_1^{(i)}$ can still be used to develop a search strategy that finds $\alpha_1^{(1)}$ and $\alpha_1^{(2)}$ at the same time. An example of a search strategy for problems with multiple linear penalties is given in Ref.~\cite{experimentalpaper}.

\begin{figure}
    \centering
    \includegraphics[width=\columnwidth]{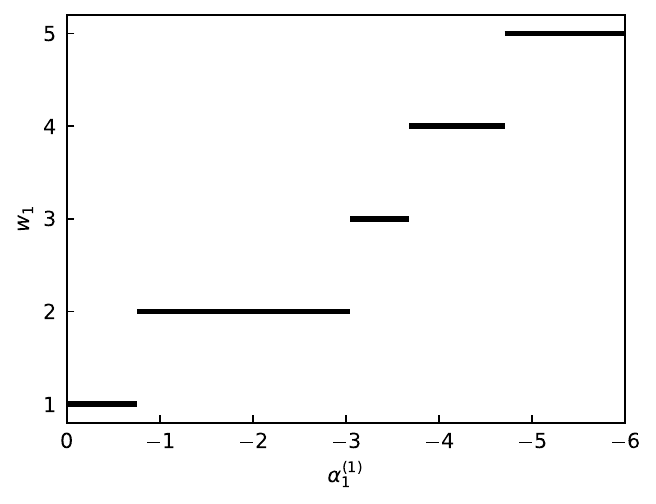}
    \caption{Ground state Hamming weight $w_1$ of the first quarter of a two-quarter promotion cannibalization problem instance plotted against the linear penalty strength $\alpha_1^{(1)}$ that is applied to the first quarter. A quadratic penalty is applied to the second quarter that constrains its ground state Hamming weight to $w_2 = 3$. The $C$ matrix is of the instance with ID \texttt{8\_19}, which we did not find any combination of linear penalty strengths for which applying linear penalties to both quarters was successful in creating a ground state in $H_P$ that satisfies $w_1 = w_2 = 3$. Note that the penalty strength $\alpha_1^{(1)}$ is decreasing along the x-axis.
    }
    \label{fig:linear_quadratic_penalties_hamming_weight}
\end{figure}

In situations where applying multiple linear penalties is not able to implement the desired constraints, it is often possible to resolve this issue by switching some of the linear penalty functions to quadratic penalties. In many cases, this maintains some of the advantages of using the linear penalty method. An example where this approach can be used is the instance shown in Fig.~\ref{fig:linear_penalties_colour_map}(b), for which we did not find any combination of $\alpha_1^{(1)}$ and $\alpha_1^{(2)}$ values that satisfies both constraints in Eq.~\eqref{eq:constraint_C1} with $A = 3$. Since our search was conducted at a finite precision, this implies that the region in which these constraints are satisfied either does not exist or is very small. In Fig.~\ref{fig:linear_quadratic_penalties_hamming_weight}, we consider the same problem instance and instead apply a linear penalty to the first quarter and a quadratic penalty to the second quarter. We find that with this scheme, there exists an interval of values of $\alpha_1^{(1)}$ for which all of the problem's constraints are satisfied. In Ref.~\cite{experimentalpaper}, there is a more in-depth analysis of combining linear and quadratic penalties using a four-quarter variation of the promotion cannibalization problem. An alternative approach for cases where linear penalties are not successful in implementing all constraints is proposed in Ref.~\cite{Kuramata2021}, where post-processing of infeasible solutions is used to obtain feasible solutions.

\section{\label{sec:simulation_results}Performance in simulation}

We have simulated QA and the QAOA in solving the single-quarter promotion cannibalization problem. Both sets of simulations are of closed-system dynamics and they assume all-to-all connectivity of the physical qubits. The constraint that we consider in these simulations is to select $A=3$ products to promote. We used the quadratic penalty strength $\alpha_2=2$ for all instances. We chose this value as it produced both a good success probability and feasible probability for most instances at the range of problem sizes we've considered. This is supported by Fig.~\ref{fig:qa_quadratic_penalty_strength} and similar figures in Appendix~\ref{app:quadratic_penalty_strength_choice}. The linear penalty strengths $\alpha_1$ were selected uniformly at random from the interval of values that implement the $A=3$ constraint in the ground state of $H_P$ for each instance. In some cases, no such interval could be found up to a precision of $10^{-5}$, either because the interval does not exist or because it is smaller than the precision of the search. For these instances, a simulation using the linear penalty method was not performed. We have not performed an analysis of the time required to find good values of $\alpha_1$ in this section. Therefore, it is important to note that this would result in an overhead when using the linear penalty method in practice.

In real-world scenarios, $C$ matrices are sparse. This would result in $H_P$ having fewer non-zero couplings when using linear penalties than when using quadratic penalties, which could contribute to differences in performance, as discussed in Sec.~\ref{sec:background}. For example, a real-world problem might be concerned with 1,000 products that each have nonzero cannibalization interactions with 5 other products on average. In this example, switching from using the quadratic penalty method to the linear penalty method would result in a $\approx 99.5 \%$ reduction in the number of nonzero couplings in the objective function. This is because the linear penalty method maintains the sparsity of the $C$ matrix in the objective function, whereas the quadratic penalty method introduces nonzero couplings between all pairs of variables in the single-quarter problem. Due to the small problem sizes that we are limited to in simulation, all $C$ instances used in this work are fully connected. Therefore, our numerical results do not reflect the effect that the sparsity of $C$ matrices would have on the performance of the linear penalty method at large problem sizes.

\subsection{Quantum annealing simulation}

\begin{figure}
    \centering
    \includegraphics[width=\columnwidth]{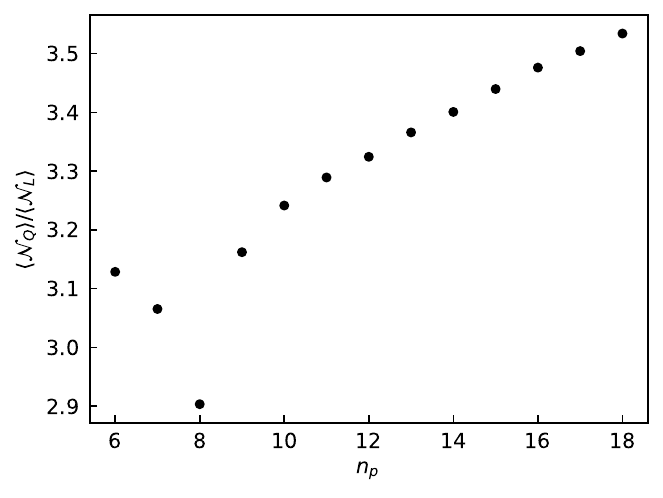}
    \caption{Ratio between the average normalisation factor of the Ising Hamiltonian when using the quadratic penalty method, $\mathcal{N}_Q$, and when using the linear penalty method, $\mathcal{N}_L$, plotted against the number of products $n_p$ for instances of the single-quarter promotion cannibalization problem.
    }
    \label{fig:normalisation_factor_ratios}
\end{figure}

In our simulations of QA, the Ising Hamiltonians were normalised by a factor $\mathcal{N}$, as in Eq.~\eqref{eq:normalised_ising_hamiltonian}, such that the maximum coupling strength is 1 and the maximum local field strength is 3. This reflects the finite energy scales that are realisable on physical devices. When using the linear penalty method, the normalisation factor $\mathcal{N}_L$ of a given problem instance is different to the normalisation factor $\mathcal{N}_Q$ when using the quadratic penalty method. As mentioned in Sec.~\ref{sec:background}, the linear penalty method often produces a smaller normalisation factor, which benefits the dynamic range of qubit interactions. In Fig.~\ref{fig:normalisation_factor_ratios}, we see that for all of the problem sizes we've considered, $\mathcal{N}_Q$ is significantly larger than $\mathcal{N}_L$ on average. The larger effective dynamic range when using the linear penalty method can lead to a more efficient exploration of the search space in QA.

\begin{figure}
    \centering
    \includegraphics[width=\columnwidth]{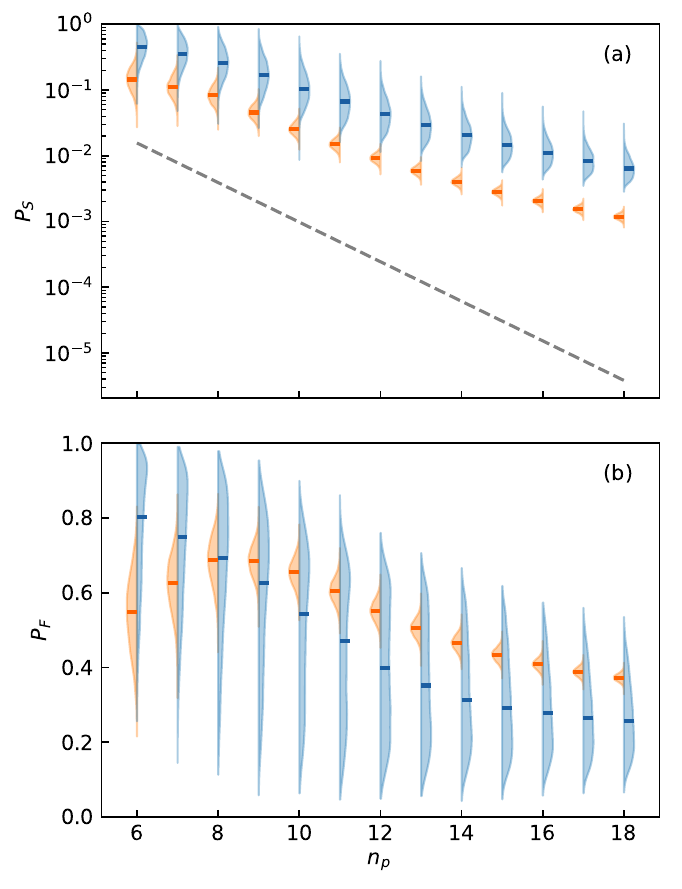}
    \caption{(a) The success probability $P_S$ of coherent quantum annealing using the quadratic (orange) and linear (blue) penalty methods after an anneal time of $t=10$ against the number of products $n_p$ in a single-quarter promotion cannibalization problem. The distribution of $P_S$ for the different problem instances is plotted alongside a bar representing the median value of the distribution. Note that the y-axis scale and histogram bin widths are logarithmic. (b) Similar plot for the probability $P_F$ of measuring a feasible state on axes with linear scales.
    }
    \label{fig:qa_sim_results}
\end{figure}

For the range of number of products we have considered in the QA simulations, the linear penalty method outperforms the quadratic penalty method on average in finding the optimal solution, as shown in Fig.~\ref{fig:qa_sim_results}(a). Both methods are more successful than random guessing. As the number of products is increased, the variation in performance across different instances shrinks when using the quadratic method, whereas for the linear method it remains large.

In Fig.~\ref{fig:qa_sim_results}(b), we find that when the number of products is greater than 8, the quadratic penalty method is on average more likely to sample feasible states. The width of the distribution of feasible probabilities is much larger when using the linear method than the quadratic method. This indicates that with our parameter choices, the ability for the linear penalty method to be effective in implementing the constraint is highly dependent on the problem instance, and it explains why the success probability distributions have similarly large widths for the linear method. The fact that the linear penalty method produces a lower feasible probability than the quadratic method on some instances and a higher feasible probability on others suggests that hybrid approaches that attempt using both penalty methods would be effective in practice. 

These simulations are of the ideal regime where the effects of limited qubit connectivity and noise do not play a role. Thus, while the advantage of using fewer couplings with the linear penalty method is more clear when the hardware does not support all-to-all connectivity, these simulation results indicate that the linear penalty method may also benefit algorithms on fault-tolerant quantum devices with full qubit connectivity. This is complementary to promising experimental results obtained on D-Wave annealers~\cite{experimentalpaper, Ohzeki2020, Yu2021, Kuramata2021}, where the effects of hardware limitations are prominent.

\subsection{Quantum approximate optimization algorithm simulation}

\begin{figure}
    \centering
    \includegraphics[width=\columnwidth]{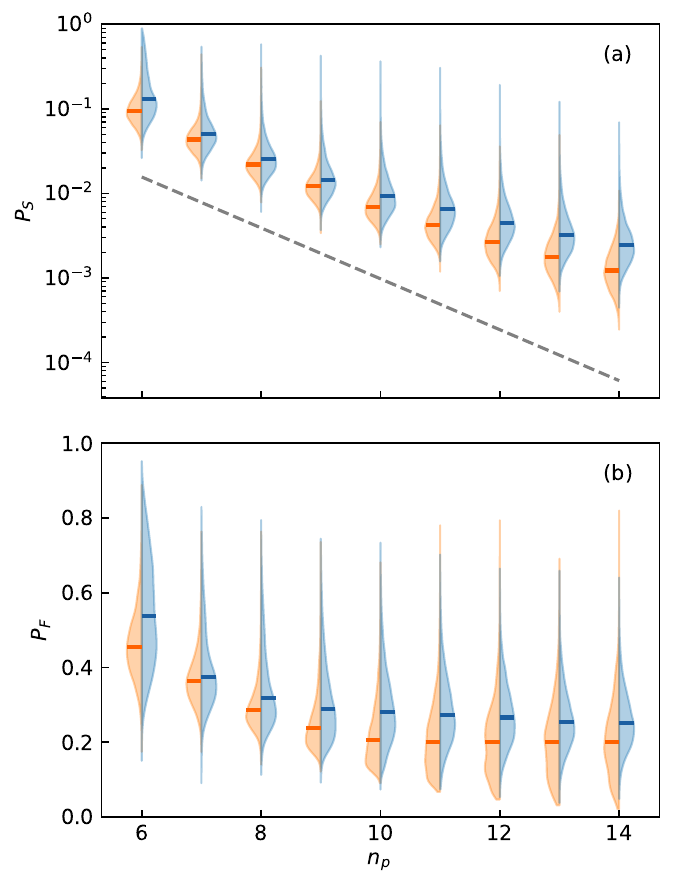}
    \caption{(a) The inferred success probability $P_S$ of the QAOA with $p=8$ layers using the quadratic (orange) and linear (blue) penalty methods against the number of products $n_p$ in a single-quarter promotion cannibalization problem. The distribution of $P_S$ for the different problem instances is plotted alongside a bar representing the median value of the distribution. Note that the y-axis scale and histogram bin widths are logarithmic. (b) Similar plot for the probability $P_F$ of measuring a feasible state on axes with linear scales.
    }
    \label{fig:qaoa_sim_results}
\end{figure}

We performed another set of simulations that are of the QAOA with $p=8$ layers of gates. The same problem instances were used as in the QA simulations. Due to time and computational resource limitations, we simulated problems with up to $n_p=14$ products. Fig.~\ref{fig:qaoa_sim_results} shows the inferred success probabilities and feasible probabilities for the QAOA plotted against the number of products in the problem. We find that the QAOA performs better than random guessing on all of the problem instances, whether the linear or quadratic penalty method is used. For all values of $n_p$ we considered, the linear penalty method produces more favourable success and feasible probabilities on average compared to the quadratic penalty method. Since $H_P$ does not need to be normalised in the QAOA, these results are encouraging as they suggest that the linear penalty method continues to have advantages even when there are no strict limitations on the energy scales of interactions.

Comparing the results from the two algorithms, we find that the difference in average success probabilities between the two penalty methods is smaller for the QAOA simulations than the QA simulations. We note that the QAOA and QA results are dependent on the choices of $p$ and $t_f$ respectively, so a fair comparison of the two algorithms would require these two parameters to be chosen in such a way that the effective anneal times are the same. Since the purpose of this study is not to compare these two algorithms, we have not done this. An interesting direction for future work would be to determine how the performance difference between the linear and quadratic penalty methods depends on the number of QAOA layers $p$ or anneal time $t_f$.

Something else that is not explored in our analysis of these simulations is the time required to execute the QAOA algorithm and how it differs between the two penalty methods. In real-world problems, where $C$ matrices are typically sparse, encoding constraints with the linear penalty method would require fewer non-zero couplings in $H_P$. Therefore, the quantum circuit to implement the unitary $e^{-i \gamma_k H_P}$ requires fewer two-qubit gates, making it more efficient to run the QAOA circuit with the linear method. Another way in which the runtime could differ between the penalty methods would be if it were easier to classically optimize the angles $\boldsymbol{\beta}$ and $\boldsymbol{\gamma}$ using one penalty method than the other, if such a difference exists.

\section{\label{sec:conclusions}Conclusions}

In this work, we have investigated the linear penalty method for encoding constraints and made two theoretical arguments for why the method can lead to better performance in quantum optimization than the quadratic penalty method, which is currently the standard approach. The first argument is based on the fact that the linear penalty method does not introduce any additional quadratic terms to the objective function, and the second is based on considerations of the energy scales associated with the two types of penalty functions. The linear penalty method is not always successful in exactly implementing a desired constraint. However, we have identified a type of customer data science problem for which it is often successful because all quadratic terms in the problem's objective function have non-negative coefficients. There may be other structures that make problems more amenable to the linear penalty method. While this is beyond the scope of our current work, it provides an interesting direction for future research.

We have studied the behaviour of the linear and quadratic penalty methods with respect to changes in their penalty strengths $\alpha_1$ and $\alpha_2$. The linear penalty method is more sensitive to its penalty strength, making it necessary to search for different values of $\alpha_1$ for each problem instance. While this is an extra step that often isn't necessary when using the quadratic penalty method, the monotonic relationship between $\alpha_1$ and Hamming weight can make this search easy to perform in practice. In problems where multiple linear penalties are applied, the different linear penalty strengths can influence each other, making the search for $\alpha_1$ more difficult. For cases where one or more constraints cannot be successfully implemented with linear penalties, we have shown that it is sometimes possible to use a combination of quadratic and linear penalties to implement the desired constraints.

Our simulations of QA and the QAOA indicate that there may be a performance enhancement when using the linear penalty method over the quadratic method. We suspect that the linear penalty method would more convincingly outperform at larger problem sizes, where the dynamic range effects are more prominent and the problems may be more sparse. This is supported by the findings in Ref.~\cite{experimentalpaper}, where larger instances of customer data science problems are tackled on real quantum hardware.

This work only considered the application of linear penalties to equality constraints, but they can also be used for inequality constraints~\cite{Yu2021}. In this case, the linear penalty would create an approximation of the original problem as it would penalise different feasible solutions by different amounts. However, the larger range of acceptable Hamming weights may make it easier to find values of the linear penalty strengths that produce feasible solutions than for equality constraints. Further research is required to investigate this.

\vspace{5mm}

The data and code that support the findings of this study are openly available at Ref.~\cite{Mirkarimi2024data}.

\begin{acknowledgments}

We thank Pim van den Heuvel for providing constructive feedback. NC was supported by UK Engineering and Physical Sciences Research Council (EPSRC) grant number EP/T026715/2 (CCP-QC). PM was supported by UK EPSRC Doctoral Training Funds awarded to Durham University (EP/T518001/1) in partnership with dunnhumby.

\end{acknowledgments}

\appendix

\section{\label{app:quadratic_penalty_strength_choice}Quadratic penalty strength choice for numerical simulations}

\begin{figure*}
    \centering
    \includegraphics[width=\textwidth]{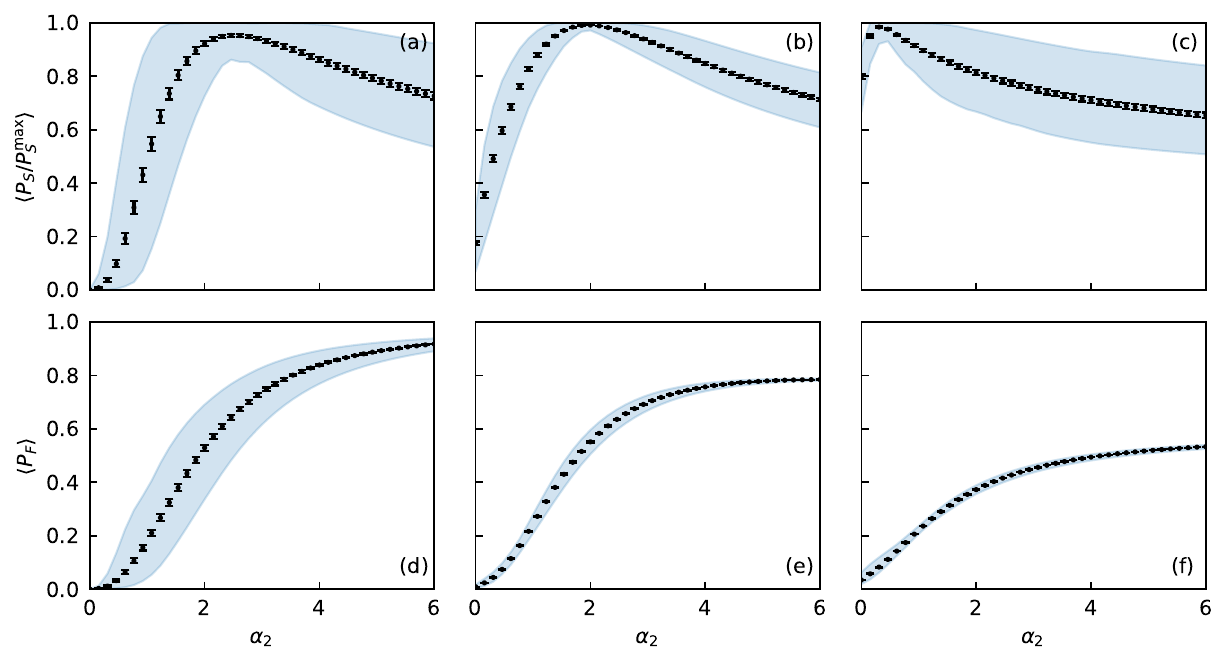}
    \caption{We have measured the probability $P_S$ after an anneal time $t_f=10$ of measuring the optimal state normalised by an estimate of the maximum success probability $P_S^\mathrm{max}$ for closed-system QA at different values of the penalty parameter $\alpha_2$. We plot the average of $P_S/P_S^\mathrm{max}$ for 100 instances of the promotion cannibalization problem with (a) 6, (b) 12, and (c) 18 products. Error bars represent the standard error in the mean. The blue shaded regions contain the 5th to 95th percentile values of $P_S/P_S^\mathrm{max}$. Note that $P_S^\mathrm{max}$ is calculated for each instance separately and is estimated by taking the maximum of $P_S$ over the plotted values of $\alpha_2$. We similarly plot the probability $P_F$ of measuring a feasible state averaged over the same 100 instances with (d) 6, (e) 12, and (f) 18 products.}
    \label{fig:qa_quadratic_penalty_strengths}
\end{figure*}

For our simulations of QA and the QAOA, we used the quadratic penalty strength $\alpha_2 = 2$. To choose this value, we measured the success probability $P_S$ and feasible probability $P_F$ of QA over a range of values of $\alpha_2$. The resulting plots are shown in Fig.~\ref{fig:qa_quadratic_penalty_strengths} for three different numbers of products $n_p$. Note that Fig.~\ref{fig:qa_quadratic_penalty_strengths}(b) and Fig.~\ref{fig:qa_quadratic_penalty_strengths}(e) show the same data as in Fig.~\ref{fig:qa_quadratic_penalty_strength}. As $n_p$ is increased, we find that $P_S$ peaks at smaller values of $\alpha_2$ on average, whereas the shape of the curve for $P_F$ does not change as significantly. There is a gradual drop in $P_S$ as $\alpha_2$ is increased beyond the value that maximises $P_S$. Across the range of values of $n_p$ we've considered in our simulations, setting $\alpha_2 = 2$ results in a good compromise between maximising $P_S$ and maximising $P_F$ for most problem instances.

\begin{figure}
    \centering
    \includegraphics[width=\columnwidth]{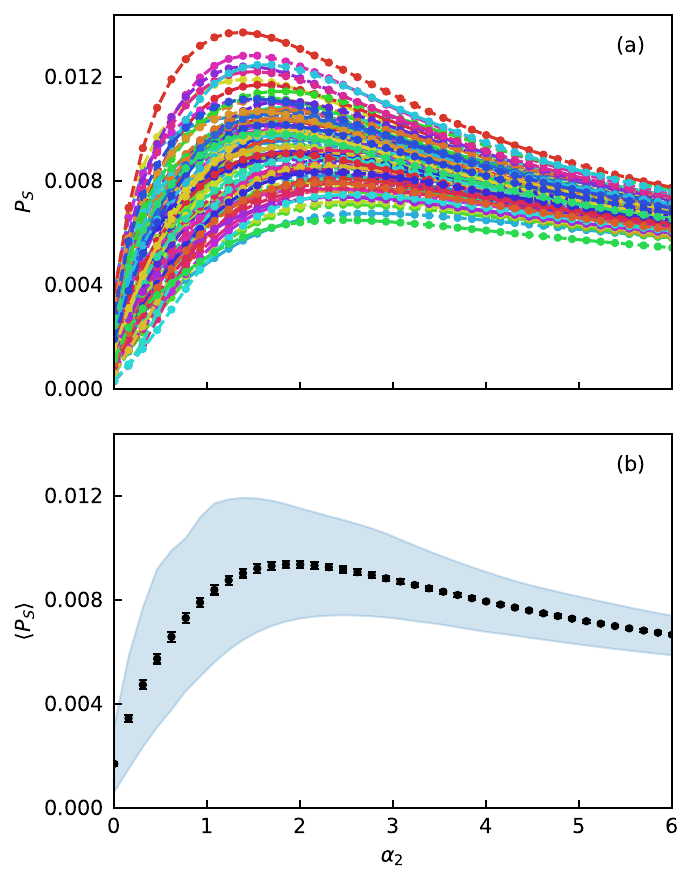}
    \caption{(a) Success probability $P_S$ of QA plotted against the quadratic penalty strength $\alpha_2$ for the single-quarter cannibalization problem with $n_p=12$ products. This is plotted for 100 instances of the problem that are represented by different colours, with dashed lines connecting each instance's points to guide the eye. (b) $P_S$ is averaged over the different problem instances and and plotted with error bars representing the standard error in the mean. The blue shaded region shows the area that contains the 5th to 95th percentile values of $P_S$.
    }
    \label{fig:qa_quadratic_penalty_strength_unnormalised}
\end{figure}

Fig.~\ref{fig:qa_quadratic_penalty_strength} and Fig.~\ref{fig:qa_quadratic_penalty_strengths} plot average normalised success probabilities of QA. The normalisation was performed in order to avoid the averages being skewed in favour of instances for which the maximum success probability is larger. In Fig.~\ref{fig:qa_quadratic_penalty_strength_unnormalised}, we plot unnormalised success probabilities for problem instances with $n_p = 12$ products. Comparing this with Fig.~\ref{fig:qa_quadratic_penalty_strength}, we see that the average success probability peaks at roughly the same value of $\alpha_2$ regardless of whether a normalisation is performed.

\begin{thebibliography}{62}%
\makeatletter
\providecommand \@ifxundefined [1]{%
 \@ifx{#1\undefined}
}%
\providecommand \@ifnum [1]{%
 \ifnum #1\expandafter \@firstoftwo
 \else \expandafter \@secondoftwo
 \fi
}%
\providecommand \@ifx [1]{%
 \ifx #1\expandafter \@firstoftwo
 \else \expandafter \@secondoftwo
 \fi
}%
\providecommand \natexlab [1]{#1}%
\providecommand \enquote  [1]{``#1''}%
\providecommand \bibnamefont  [1]{#1}%
\providecommand \bibfnamefont [1]{#1}%
\providecommand \citenamefont [1]{#1}%
\providecommand \href@noop [0]{\@secondoftwo}%
\providecommand \href [0]{\begingroup \@sanitize@url \@href}%
\providecommand \@href[1]{\@@startlink{#1}\@@href}%
\providecommand \@@href[1]{\endgroup#1\@@endlink}%
\providecommand \@sanitize@url [0]{\catcode `\\12\catcode `\$12\catcode `\&12\catcode `\#12\catcode `\^12\catcode `\_12\catcode `\%12\relax}%
\providecommand \@@startlink[1]{}%
\providecommand \@@endlink[0]{}%
\providecommand \url  [0]{\begingroup\@sanitize@url \@url }%
\providecommand \@url [1]{\endgroup\@href {#1}{\urlprefix }}%
\providecommand \urlprefix  [0]{URL }%
\providecommand \Eprint [0]{\href }%
\providecommand \doibase [0]{https://doi.org/}%
\providecommand \selectlanguage [0]{\@gobble}%
\providecommand \bibinfo  [0]{\@secondoftwo}%
\providecommand \bibfield  [0]{\@secondoftwo}%
\providecommand \translation [1]{[#1]}%
\providecommand \BibitemOpen [0]{}%
\providecommand \bibitemStop [0]{}%
\providecommand \bibitemNoStop [0]{.\EOS\space}%
\providecommand \EOS [0]{\spacefactor3000\relax}%
\providecommand \BibitemShut  [1]{\csname bibitem#1\endcsname}%
\let\auto@bib@innerbib\@empty
\bibitem [{\citenamefont {Kadowaki}\ and\ \citenamefont {Nishimori}(1998)}]{Kadowaki1998}%
  \BibitemOpen
  \bibfield  {author} {\bibinfo {author} {\bibfnamefont {T.}~\bibnamefont {Kadowaki}}\ and\ \bibinfo {author} {\bibfnamefont {H.}~\bibnamefont {Nishimori}},\ }\bibfield  {title} {\bibinfo {title} {{Quantum annealing in the transverse Ising model}},\ }\href {https://doi.org/10.1103/PhysRevE.58.5355} {\bibfield  {journal} {\bibinfo  {journal} {Physical Review E}\ }\textbf {\bibinfo {volume} {58}},\ \bibinfo {pages} {5355} (\bibinfo {year} {1998})}\BibitemShut {NoStop}%
\bibitem [{\citenamefont {Brooke}\ \emph {et~al.}(1999)\citenamefont {Brooke}, \citenamefont {Bitko}, \citenamefont {Rosenbaum},\ and\ \citenamefont {Aeppli}}]{Brooke1999}%
  \BibitemOpen
  \bibfield  {author} {\bibinfo {author} {\bibfnamefont {J.}~\bibnamefont {Brooke}}, \bibinfo {author} {\bibfnamefont {D.}~\bibnamefont {Bitko}}, \bibinfo {author} {\bibfnamefont {T.~F.}\ \bibnamefont {Rosenbaum}},\ and\ \bibinfo {author} {\bibfnamefont {G.}~\bibnamefont {Aeppli}},\ }\bibfield  {title} {\bibinfo {title} {{Quantum annealing of a disordered magnet}},\ }\href {https://doi.org/10.1126/science.284.5415.779} {\bibfield  {journal} {\bibinfo  {journal} {Science}\ }\textbf {\bibinfo {volume} {284}},\ \bibinfo {pages} {779} (\bibinfo {year} {1999})}\BibitemShut {NoStop}%
\bibitem [{\citenamefont {Farhi}\ \emph {et~al.}(2001)\citenamefont {Farhi}, \citenamefont {Goldstone}, \citenamefont {Gutmann}, \citenamefont {Lapan}, \citenamefont {Lundgren},\ and\ \citenamefont {Preda}}]{Farhi2001}%
  \BibitemOpen
  \bibfield  {author} {\bibinfo {author} {\bibfnamefont {E.}~\bibnamefont {Farhi}}, \bibinfo {author} {\bibfnamefont {J.}~\bibnamefont {Goldstone}}, \bibinfo {author} {\bibfnamefont {S.}~\bibnamefont {Gutmann}}, \bibinfo {author} {\bibfnamefont {J.}~\bibnamefont {Lapan}}, \bibinfo {author} {\bibfnamefont {A.}~\bibnamefont {Lundgren}},\ and\ \bibinfo {author} {\bibfnamefont {D.}~\bibnamefont {Preda}},\ }\bibfield  {title} {\bibinfo {title} {{A Quantum Adiabatic Evolution Algorithm Applied to Random Instances of an NP-Complete Problem}},\ }\href {https://doi.org/10.1126/science.1057726} {\bibfield  {journal} {\bibinfo  {journal} {Science}\ }\textbf {\bibinfo {volume} {292}},\ \bibinfo {pages} {472} (\bibinfo {year} {2001})}\BibitemShut {NoStop}%
\bibitem [{\citenamefont {Farhi}\ \emph {et~al.}(2014)\citenamefont {Farhi}, \citenamefont {Goldstone},\ and\ \citenamefont {Gutmann}}]{Farhi2014}%
  \BibitemOpen
  \bibfield  {author} {\bibinfo {author} {\bibfnamefont {E.}~\bibnamefont {Farhi}}, \bibinfo {author} {\bibfnamefont {J.}~\bibnamefont {Goldstone}},\ and\ \bibinfo {author} {\bibfnamefont {S.}~\bibnamefont {Gutmann}},\ }\href {https://doi.org/10.48550/arXiv.1411.4028} {\bibinfo {title} {{A Quantum Approximate Optimization Algorithm}}} (\bibinfo {year} {2014}),\ \bibinfo {note} {arXiv preprint arXiv:1411.4028}\BibitemShut {NoStop}%
\bibitem [{\citenamefont {Brady}\ and\ \citenamefont {Hadfield}(2023)}]{Brady2023}%
  \BibitemOpen
  \bibfield  {author} {\bibinfo {author} {\bibfnamefont {L.~T.}\ \bibnamefont {Brady}}\ and\ \bibinfo {author} {\bibfnamefont {S.}~\bibnamefont {Hadfield}},\ }\href {https://doi.org/10.48550/arXiv.2309.13110} {\bibinfo {title} {{Iterative Quantum Algorithms for Maximum Independent Set: A Tale of Low-Depth Quantum Algorithms}}} (\bibinfo {year} {2023}),\ \bibinfo {note} {arXiv preprint arXiv:2309.13110}\BibitemShut {NoStop}%
\bibitem [{\citenamefont {King}\ \emph {et~al.}(2024)\citenamefont {King} \emph {et~al.}}]{King2024}%
  \BibitemOpen
  \bibfield  {author} {\bibinfo {author} {\bibfnamefont {A.~D.}\ \bibnamefont {King}} \emph {et~al.},\ }\href {http://arxiv.org/abs/2403.00910} {\bibinfo {title} {{Computational supremacy in quantum simulation}}} (\bibinfo {year} {2024}),\ \bibinfo {note} {arXiv preprint arXiv:2403.00910}\BibitemShut {NoStop}%
\bibitem [{\citenamefont {Or{\'{u}}s}\ \emph {et~al.}(2019)\citenamefont {Or{\'{u}}s}, \citenamefont {Mugel},\ and\ \citenamefont {Lizaso}}]{Orus2019}%
  \BibitemOpen
  \bibfield  {author} {\bibinfo {author} {\bibfnamefont {R.}~\bibnamefont {Or{\'{u}}s}}, \bibinfo {author} {\bibfnamefont {S.}~\bibnamefont {Mugel}},\ and\ \bibinfo {author} {\bibfnamefont {E.}~\bibnamefont {Lizaso}},\ }\bibfield  {title} {\bibinfo {title} {{Quantum computing for finance: Overview and prospects}},\ }\href {https://doi.org/10.1016/j.revip.2019.100028} {\bibfield  {journal} {\bibinfo  {journal} {Reviews in Physics}\ }\textbf {\bibinfo {volume} {4}},\ \bibinfo {pages} {100028} (\bibinfo {year} {2019})}\BibitemShut {NoStop}%
\bibitem [{\citenamefont {Venturelli}\ and\ \citenamefont {Kondratyev}(2019)}]{Venturelli2019}%
  \BibitemOpen
  \bibfield  {author} {\bibinfo {author} {\bibfnamefont {D.}~\bibnamefont {Venturelli}}\ and\ \bibinfo {author} {\bibfnamefont {A.}~\bibnamefont {Kondratyev}},\ }\bibfield  {title} {\bibinfo {title} {{Reverse quantum annealing approach to portfolio optimization problems}},\ }\href {https://doi.org/10.1007/s42484-019-00001-w} {\bibfield  {journal} {\bibinfo  {journal} {Quantum Machine Intelligence}\ }\textbf {\bibinfo {volume} {1}},\ \bibinfo {pages} {17} (\bibinfo {year} {2019})}\BibitemShut {NoStop}%
\bibitem [{\citenamefont {Kitai}\ \emph {et~al.}(2020)\citenamefont {Kitai}, \citenamefont {Guo}, \citenamefont {Ju}, \citenamefont {Tanaka}, \citenamefont {Tsuda}, \citenamefont {Shiomi},\ and\ \citenamefont {Tamura}}]{Kitai2020}%
  \BibitemOpen
  \bibfield  {author} {\bibinfo {author} {\bibfnamefont {K.}~\bibnamefont {Kitai}}, \bibinfo {author} {\bibfnamefont {J.}~\bibnamefont {Guo}}, \bibinfo {author} {\bibfnamefont {S.}~\bibnamefont {Ju}}, \bibinfo {author} {\bibfnamefont {S.}~\bibnamefont {Tanaka}}, \bibinfo {author} {\bibfnamefont {K.}~\bibnamefont {Tsuda}}, \bibinfo {author} {\bibfnamefont {J.}~\bibnamefont {Shiomi}},\ and\ \bibinfo {author} {\bibfnamefont {R.}~\bibnamefont {Tamura}},\ }\bibfield  {title} {\bibinfo {title} {{Designing metamaterials with quantum annealing and factorization machines}},\ }\href {https://doi.org/10.1103/PhysRevResearch.2.013319} {\bibfield  {journal} {\bibinfo  {journal} {Physical Review Research}\ }\textbf {\bibinfo {volume} {2}},\ \bibinfo {pages} {013319} (\bibinfo {year} {2020})}\BibitemShut {NoStop}%
\bibitem [{\citenamefont {Stollenwerk}\ \emph {et~al.}(2020)\citenamefont {Stollenwerk}, \citenamefont {O'Gorman}, \citenamefont {Venturelli}, \citenamefont {Mandra}, \citenamefont {Rodionova}, \citenamefont {Ng}, \citenamefont {Sridhar}, \citenamefont {Rieffel},\ and\ \citenamefont {Biswas}}]{Stollenwerk2020}%
  \BibitemOpen
  \bibfield  {author} {\bibinfo {author} {\bibfnamefont {T.}~\bibnamefont {Stollenwerk}}, \bibinfo {author} {\bibfnamefont {B.}~\bibnamefont {O'Gorman}}, \bibinfo {author} {\bibfnamefont {D.}~\bibnamefont {Venturelli}}, \bibinfo {author} {\bibfnamefont {S.}~\bibnamefont {Mandra}}, \bibinfo {author} {\bibfnamefont {O.}~\bibnamefont {Rodionova}}, \bibinfo {author} {\bibfnamefont {H.}~\bibnamefont {Ng}}, \bibinfo {author} {\bibfnamefont {B.}~\bibnamefont {Sridhar}}, \bibinfo {author} {\bibfnamefont {E.~G.}\ \bibnamefont {Rieffel}},\ and\ \bibinfo {author} {\bibfnamefont {R.}~\bibnamefont {Biswas}},\ }\bibfield  {title} {\bibinfo {title} {{Quantum Annealing Applied to De-Conflicting Optimal Trajectories for Air Traffic Management}},\ }\href {https://doi.org/10.1109/TITS.2019.2891235} {\bibfield  {journal} {\bibinfo  {journal} {IEEE Transactions on Intelligent Transportation Systems}\ }\textbf {\bibinfo {volume} {21}},\ \bibinfo {pages} {285} (\bibinfo {year} {2020})}\BibitemShut {NoStop}%
\bibitem [{\citenamefont {Fox}\ \emph {et~al.}(2021)\citenamefont {Fox}, \citenamefont {Branson},\ and\ \citenamefont {Walker}}]{Fox2021}%
  \BibitemOpen
  \bibfield  {author} {\bibinfo {author} {\bibfnamefont {D.~M.}\ \bibnamefont {Fox}}, \bibinfo {author} {\bibfnamefont {K.~M.}\ \bibnamefont {Branson}},\ and\ \bibinfo {author} {\bibfnamefont {R.~C.}\ \bibnamefont {Walker}},\ }\bibfield  {title} {\bibinfo {title} {{mRNA codon optimization with quantum computers}},\ }\href {https://doi.org/10.1371/journal.pone.0259101} {\bibfield  {journal} {\bibinfo  {journal} {PLOS ONE}\ }\textbf {\bibinfo {volume} {16}},\ \bibinfo {pages} {1} (\bibinfo {year} {2021})}\BibitemShut {NoStop}%
\bibitem [{\citenamefont {Yarkoni}\ \emph {et~al.}(2022)\citenamefont {Yarkoni}, \citenamefont {Raponi}, \citenamefont {B{\"{a}}ck},\ and\ \citenamefont {Schmitt}}]{Yarkoni2021}%
  \BibitemOpen
  \bibfield  {author} {\bibinfo {author} {\bibfnamefont {S.}~\bibnamefont {Yarkoni}}, \bibinfo {author} {\bibfnamefont {E.}~\bibnamefont {Raponi}}, \bibinfo {author} {\bibfnamefont {T.}~\bibnamefont {B{\"{a}}ck}},\ and\ \bibinfo {author} {\bibfnamefont {S.}~\bibnamefont {Schmitt}},\ }\bibfield  {title} {\bibinfo {title} {{Quantum annealing for industry applications: introduction and review}},\ }\href {https://doi.org/10.1088/1361-6633/ac8c54} {\bibfield  {journal} {\bibinfo  {journal} {Reports on Progress in Physics}\ }\textbf {\bibinfo {volume} {85}},\ \bibinfo {pages} {104001} (\bibinfo {year} {2022})}\BibitemShut {NoStop}%
\bibitem [{\citenamefont {Nishimura}\ \emph {et~al.}(2019)\citenamefont {Nishimura}, \citenamefont {Tanahashi}, \citenamefont {Suganuma}, \citenamefont {Miyama},\ and\ \citenamefont {Ohzeki}}]{Nishimura2019}%
  \BibitemOpen
  \bibfield  {author} {\bibinfo {author} {\bibfnamefont {N.}~\bibnamefont {Nishimura}}, \bibinfo {author} {\bibfnamefont {K.}~\bibnamefont {Tanahashi}}, \bibinfo {author} {\bibfnamefont {K.}~\bibnamefont {Suganuma}}, \bibinfo {author} {\bibfnamefont {M.~J.}\ \bibnamefont {Miyama}},\ and\ \bibinfo {author} {\bibfnamefont {M.}~\bibnamefont {Ohzeki}},\ }\bibfield  {title} {\bibinfo {title} {{Item Listing Optimization for E-Commerce Websites Based on Diversity}},\ }\bibfield  {journal} {\bibinfo  {journal} {Frontiers in Computer Science}\ }\textbf {\bibinfo {volume} {1}},\ \href {https://doi.org/10.3389/fcomp.2019.00002} {10.3389/fcomp.2019.00002} (\bibinfo {year} {2019})\BibitemShut {NoStop}%
\bibitem [{\citenamefont {Weinberg}\ \emph {et~al.}(2023)\citenamefont {Weinberg}, \citenamefont {Sanches}, \citenamefont {Ide}, \citenamefont {Kamiya},\ and\ \citenamefont {Correll}}]{Weinberg2023}%
  \BibitemOpen
  \bibfield  {author} {\bibinfo {author} {\bibfnamefont {S.~J.}\ \bibnamefont {Weinberg}}, \bibinfo {author} {\bibfnamefont {F.}~\bibnamefont {Sanches}}, \bibinfo {author} {\bibfnamefont {T.}~\bibnamefont {Ide}}, \bibinfo {author} {\bibfnamefont {K.}~\bibnamefont {Kamiya}},\ and\ \bibinfo {author} {\bibfnamefont {R.}~\bibnamefont {Correll}},\ }\bibfield  {title} {\bibinfo {title} {{Supply chain logistics with quantum and classical annealing algorithms}},\ }\href {https://doi.org/10.1038/s41598-023-31765-8} {\bibfield  {journal} {\bibinfo  {journal} {Scientific Reports}\ }\textbf {\bibinfo {volume} {13}},\ \bibinfo {pages} {4770} (\bibinfo {year} {2023})}\BibitemShut {NoStop}%
\bibitem [{\citenamefont {Czerniachowska}(2022)}]{Czerniachowska2022}%
  \BibitemOpen
  \bibfield  {author} {\bibinfo {author} {\bibfnamefont {K.}~\bibnamefont {Czerniachowska}},\ }\bibfield  {title} {\bibinfo {title} {{A genetic algorithm for the retail shelf space allocation problem with virtual segments}},\ }\href {https://doi.org/10.1007/s12597-021-00551-3} {\bibfield  {journal} {\bibinfo  {journal} {OPSEARCH}\ }\textbf {\bibinfo {volume} {59}},\ \bibinfo {pages} {364} (\bibinfo {year} {2022})}\BibitemShut {NoStop}%
\bibitem [{\citenamefont {Subramanian}\ and\ \citenamefont {Sherali}(2010)}]{Subramanian2010}%
  \BibitemOpen
  \bibfield  {author} {\bibinfo {author} {\bibfnamefont {S.}~\bibnamefont {Subramanian}}\ and\ \bibinfo {author} {\bibfnamefont {H.~D.}\ \bibnamefont {Sherali}},\ }\bibfield  {title} {\bibinfo {title} {{A fractional programming approach for retail category price optimization}},\ }\href {https://doi.org/10.1007/s10898-009-9491-2} {\bibfield  {journal} {\bibinfo  {journal} {Journal of Global Optimization}\ }\textbf {\bibinfo {volume} {48}},\ \bibinfo {pages} {263} (\bibinfo {year} {2010})}\BibitemShut {NoStop}%
\bibitem [{\citenamefont {Lucas}(2014)}]{Lucas2014}%
  \BibitemOpen
  \bibfield  {author} {\bibinfo {author} {\bibfnamefont {A.}~\bibnamefont {Lucas}},\ }\bibfield  {title} {\bibinfo {title} {{Ising formulations of many NP problems}},\ }\bibfield  {journal} {\bibinfo  {journal} {Frontiers in Physics}\ }\textbf {\bibinfo {volume} {2}},\ \href {https://doi.org/10.3389/fphy.2014.00005} {10.3389/fphy.2014.00005} (\bibinfo {year} {2014})\BibitemShut {NoStop}%
\bibitem [{\citenamefont {Lodewijks}(2019)}]{Lodewijks2019}%
  \BibitemOpen
  \bibfield  {author} {\bibinfo {author} {\bibfnamefont {B.}~\bibnamefont {Lodewijks}},\ }\href {https://doi.org/10.48550/arXiv.1911.08043} {\bibinfo {title} {{Mapping NP-hard and NP-complete optimisation problems to Quadratic Unconstrained Binary Optimisation problems}}} (\bibinfo {year} {2019}),\ \bibinfo {note} {arXiv preprint arXiv:1911.08043}\BibitemShut {NoStop}%
\bibitem [{\citenamefont {Preskill}(2018)}]{Preskill2018}%
  \BibitemOpen
  \bibfield  {author} {\bibinfo {author} {\bibfnamefont {J.}~\bibnamefont {Preskill}},\ }\bibfield  {title} {\bibinfo {title} {{Quantum Computing in the NISQ era and beyond}},\ }\href {https://doi.org/10.22331/q-2018-08-06-79} {\bibfield  {journal} {\bibinfo  {journal} {Quantum}\ }\textbf {\bibinfo {volume} {2}},\ \bibinfo {pages} {79} (\bibinfo {year} {2018})}\BibitemShut {NoStop}%
\bibitem [{\citenamefont {Choi}(2008)}]{Choi2008}%
  \BibitemOpen
  \bibfield  {author} {\bibinfo {author} {\bibfnamefont {V.}~\bibnamefont {Choi}},\ }\bibfield  {title} {\bibinfo {title} {{Minor-embedding in adiabatic quantum computation: I. The parameter setting problem}},\ }\href {https://doi.org/10.1007/s11128-008-0082-9} {\bibfield  {journal} {\bibinfo  {journal} {Quantum Information Processing}\ }\textbf {\bibinfo {volume} {7}},\ \bibinfo {pages} {193} (\bibinfo {year} {2008})}\BibitemShut {NoStop}%
\bibitem [{\citenamefont {Lechner}\ \emph {et~al.}(2015)\citenamefont {Lechner}, \citenamefont {Hauke},\ and\ \citenamefont {Zoller}}]{Lechner2015}%
  \BibitemOpen
  \bibfield  {author} {\bibinfo {author} {\bibfnamefont {W.}~\bibnamefont {Lechner}}, \bibinfo {author} {\bibfnamefont {P.}~\bibnamefont {Hauke}},\ and\ \bibinfo {author} {\bibfnamefont {P.}~\bibnamefont {Zoller}},\ }\bibfield  {title} {\bibinfo {title} {{A quantum annealing architecture with all-to-all connectivity from local interactions}},\ }\href {https://doi.org/10.1126/sciadv.1500838} {\bibfield  {journal} {\bibinfo  {journal} {Science Advances}\ }\textbf {\bibinfo {volume} {1}},\ \bibinfo {pages} {1} (\bibinfo {year} {2015})}\BibitemShut {NoStop}%
\bibitem [{\citenamefont {Rocchetto}\ \emph {et~al.}(2016)\citenamefont {Rocchetto}, \citenamefont {Benjamin},\ and\ \citenamefont {Li}}]{Rocchetto2016}%
  \BibitemOpen
  \bibfield  {author} {\bibinfo {author} {\bibfnamefont {A.}~\bibnamefont {Rocchetto}}, \bibinfo {author} {\bibfnamefont {S.~C.}\ \bibnamefont {Benjamin}},\ and\ \bibinfo {author} {\bibfnamefont {Y.}~\bibnamefont {Li}},\ }\bibfield  {title} {\bibinfo {title} {Stabilizers as a design tool for new forms of the lechner-hauke-zoller annealer},\ }\href {https://doi.org/10.1126/sciadv.1601246} {\bibfield  {journal} {\bibinfo  {journal} {Science Advances}\ }\textbf {\bibinfo {volume} {2}},\ \bibinfo {pages} {e1601246} (\bibinfo {year} {2016})}\BibitemShut {NoStop}%
\bibitem [{\citenamefont {Chancellor}(2019)}]{Chancellor2019}%
  \BibitemOpen
  \bibfield  {author} {\bibinfo {author} {\bibfnamefont {N.}~\bibnamefont {Chancellor}},\ }\bibfield  {title} {\bibinfo {title} {Domain wall encoding of discrete variables for quantum annealing and qaoa},\ }\href {https://doi.org/10.1088/2058-9565/ab33c2} {\bibfield  {journal} {\bibinfo  {journal} {Quantum Science and Technology}\ }\textbf {\bibinfo {volume} {4}},\ \bibinfo {pages} {045004} (\bibinfo {year} {2019})}\BibitemShut {NoStop}%
\bibitem [{\citenamefont {Chen}\ \emph {et~al.}(2021)\citenamefont {Chen}, \citenamefont {Stollenwerk},\ and\ \citenamefont {Chancellor}}]{Chen2021}%
  \BibitemOpen
  \bibfield  {author} {\bibinfo {author} {\bibfnamefont {J.}~\bibnamefont {Chen}}, \bibinfo {author} {\bibfnamefont {T.}~\bibnamefont {Stollenwerk}},\ and\ \bibinfo {author} {\bibfnamefont {N.}~\bibnamefont {Chancellor}},\ }\bibfield  {title} {\bibinfo {title} {Performance of domain-wall encoding for quantum annealing},\ }\href {https://doi.org/10.1109/TQE.2021.3094280} {\bibfield  {journal} {\bibinfo  {journal} {IEEE Transactions on Quantum Engineering}\ }\textbf {\bibinfo {volume} {2}},\ \bibinfo {pages} {1} (\bibinfo {year} {2021})}\BibitemShut {NoStop}%
\bibitem [{\citenamefont {Berwald}\ \emph {et~al.}(2023)\citenamefont {Berwald}, \citenamefont {Chancellor},\ and\ \citenamefont {Dridi}}]{Berwald2023}%
  \BibitemOpen
  \bibfield  {author} {\bibinfo {author} {\bibfnamefont {J.}~\bibnamefont {Berwald}}, \bibinfo {author} {\bibfnamefont {N.}~\bibnamefont {Chancellor}},\ and\ \bibinfo {author} {\bibfnamefont {R.}~\bibnamefont {Dridi}},\ }\bibfield  {title} {\bibinfo {title} {Understanding domain-wall encoding theoretically and experimentally},\ }\href {https://doi.org/10.1098/rsta.2021.0410} {\bibfield  {journal} {\bibinfo  {journal} {Philosophical Transactions of the Royal Society A: Mathematical, Physical and Engineering Sciences}\ }\textbf {\bibinfo {volume} {381}},\ \bibinfo {pages} {20210410} (\bibinfo {year} {2023})}\BibitemShut {NoStop}%
\bibitem [{\citenamefont {Ohzeki}(2020)}]{Ohzeki2020}%
  \BibitemOpen
  \bibfield  {author} {\bibinfo {author} {\bibfnamefont {M.}~\bibnamefont {Ohzeki}},\ }\bibfield  {title} {\bibinfo {title} {{Breaking limitation of quantum annealer in solving optimization problems under constraints}},\ }\href {https://doi.org/10.1038/s41598-020-60022-5} {\bibfield  {journal} {\bibinfo  {journal} {Scientific Reports}\ }\textbf {\bibinfo {volume} {10}},\ \bibinfo {pages} {3126} (\bibinfo {year} {2020})}\BibitemShut {NoStop}%
\bibitem [{\citenamefont {Willsch}\ \emph {et~al.}(2020)\citenamefont {Willsch}, \citenamefont {Willsch}, \citenamefont {{De Raedt}},\ and\ \citenamefont {Michielsen}}]{Willsch2020a}%
  \BibitemOpen
  \bibfield  {author} {\bibinfo {author} {\bibfnamefont {D.}~\bibnamefont {Willsch}}, \bibinfo {author} {\bibfnamefont {M.}~\bibnamefont {Willsch}}, \bibinfo {author} {\bibfnamefont {H.}~\bibnamefont {{De Raedt}}},\ and\ \bibinfo {author} {\bibfnamefont {K.}~\bibnamefont {Michielsen}},\ }\bibfield  {title} {\bibinfo {title} {{Support vector machines on the D-Wave quantum annealer}},\ }\href {https://doi.org/10.1016/j.cpc.2019.107006} {\bibfield  {journal} {\bibinfo  {journal} {Computer Physics Communications}\ }\textbf {\bibinfo {volume} {248}},\ \bibinfo {pages} {107006} (\bibinfo {year} {2020})}\BibitemShut {NoStop}%
\bibitem [{\citenamefont {Mirkarimi}\ \emph {et~al.}(2024{\natexlab{a}})\citenamefont {Mirkarimi}, \citenamefont {Hoyle}, \citenamefont {Williams},\ and\ \citenamefont {Chancellor}}]{experimentalpaper}%
  \BibitemOpen
  \bibfield  {author} {\bibinfo {author} {\bibfnamefont {P.}~\bibnamefont {Mirkarimi}}, \bibinfo {author} {\bibfnamefont {D.~C.}\ \bibnamefont {Hoyle}}, \bibinfo {author} {\bibfnamefont {R.}~\bibnamefont {Williams}},\ and\ \bibinfo {author} {\bibfnamefont {N.}~\bibnamefont {Chancellor}},\ }\href {https://doi.org/10.48550/arXiv.2404.05476} {\bibinfo {title} {{Experimental demonstration of improved quantum optimization with linear Ising penalties}}} (\bibinfo {year} {2024}{\natexlab{a}}),\ \bibinfo {note} {arXiv preprint arXiv:2404.05476}\BibitemShut {NoStop}%
\bibitem [{\citenamefont {Kirkpatrick}\ \emph {et~al.}(1983)\citenamefont {Kirkpatrick}, \citenamefont {Gelatt},\ and\ \citenamefont {Vecchi}}]{Kirkpatrick1983}%
  \BibitemOpen
  \bibfield  {author} {\bibinfo {author} {\bibfnamefont {S.}~\bibnamefont {Kirkpatrick}}, \bibinfo {author} {\bibfnamefont {C.~D.}\ \bibnamefont {Gelatt}},\ and\ \bibinfo {author} {\bibfnamefont {M.~P.}\ \bibnamefont {Vecchi}},\ }\bibfield  {title} {\bibinfo {title} {{Optimization by Simulated Annealing}},\ }\href {https://doi.org/10.1126/science.220.4598.671} {\bibfield  {journal} {\bibinfo  {journal} {Science}\ }\textbf {\bibinfo {volume} {220}},\ \bibinfo {pages} {671} (\bibinfo {year} {1983})}\BibitemShut {NoStop}%
\bibitem [{\citenamefont {Albash}\ and\ \citenamefont {Lidar}(2018)}]{Albash2018}%
  \BibitemOpen
  \bibfield  {author} {\bibinfo {author} {\bibfnamefont {T.}~\bibnamefont {Albash}}\ and\ \bibinfo {author} {\bibfnamefont {D.~A.}\ \bibnamefont {Lidar}},\ }\bibfield  {title} {\bibinfo {title} {{Adiabatic quantum computation}},\ }\href {https://doi.org/10.1103/RevModPhys.90.015002} {\bibfield  {journal} {\bibinfo  {journal} {Reviews of Modern Physics}\ }\textbf {\bibinfo {volume} {90}},\ \bibinfo {pages} {015002} (\bibinfo {year} {2018})}\BibitemShut {NoStop}%
\bibitem [{\citenamefont {Meredith}\ and\ \citenamefont {Maki}(2001)}]{Meredith2001}%
  \BibitemOpen
  \bibfield  {author} {\bibinfo {author} {\bibfnamefont {L.}~\bibnamefont {Meredith}}\ and\ \bibinfo {author} {\bibfnamefont {D.}~\bibnamefont {Maki}},\ }\bibfield  {title} {\bibinfo {title} {{Product cannibalization and the role of prices}},\ }\href {https://doi.org/10.1080/00036840010015769} {\bibfield  {journal} {\bibinfo  {journal} {Applied Economics}\ }\textbf {\bibinfo {volume} {33}},\ \bibinfo {pages} {1785} (\bibinfo {year} {2001})}\BibitemShut {NoStop}%
\bibitem [{\citenamefont {Aguilar-Palacios}\ \emph {et~al.}(2021)\citenamefont {Aguilar-Palacios}, \citenamefont {Munoz-Romero},\ and\ \citenamefont {Rojo-Alvarez}}]{Aguilar-Palacios2021}%
  \BibitemOpen
  \bibfield  {author} {\bibinfo {author} {\bibfnamefont {C.}~\bibnamefont {Aguilar-Palacios}}, \bibinfo {author} {\bibfnamefont {S.}~\bibnamefont {Munoz-Romero}},\ and\ \bibinfo {author} {\bibfnamefont {J.~L.}\ \bibnamefont {Rojo-Alvarez}},\ }\bibfield  {title} {\bibinfo {title} {{Causal Quantification of Cannibalization During Promotional Sales in Grocery Retail}},\ }\href {https://doi.org/10.1109/ACCESS.2021.3062222} {\bibfield  {journal} {\bibinfo  {journal} {IEEE Access}\ }\textbf {\bibinfo {volume} {9}},\ \bibinfo {pages} {34078} (\bibinfo {year} {2021})}\BibitemShut {NoStop}%
\bibitem [{\citenamefont {Nocedal}\ and\ \citenamefont {Wright}(2006)}]{nocedal1999numerical}%
  \BibitemOpen
  \bibfield  {author} {\bibinfo {author} {\bibfnamefont {J.}~\bibnamefont {Nocedal}}\ and\ \bibinfo {author} {\bibfnamefont {S.~J.}\ \bibnamefont {Wright}},\ }\href {https://doi.org/10.1007/978-0-387-40065-5} {\emph {\bibinfo {title} {{Numerical Optimization}}}},\ Springer Series in Operations Research and Financial Engineering\ (\bibinfo  {publisher} {Springer New York},\ \bibinfo {year} {2006})\BibitemShut {NoStop}%
\bibitem [{\citenamefont {{Van Thoai}}(2013)}]{VanThoai2013}%
  \BibitemOpen
  \bibfield  {author} {\bibinfo {author} {\bibfnamefont {N.}~\bibnamefont {{Van Thoai}}},\ }\bibfield  {title} {\bibinfo {title} {{Solution Methods for General Quadratic Programming Problem with Continuous and Binary Variables: Overview}},\ }in\ \href {https://doi.org/10.1007/978-3-319-00293-4_1} {\emph {\bibinfo {booktitle} {Advanced Computational Methods for Knowledge Engineering}}},\ \bibinfo {editor} {edited by\ \bibinfo {editor} {\bibfnamefont {N.~T.}\ \bibnamefont {Nguyen}}, \bibinfo {editor} {\bibfnamefont {T.}~\bibnamefont {van Do}},\ and\ \bibinfo {editor} {\bibfnamefont {H.~A.}\ \bibnamefont {le~Thi}}}\ (\bibinfo  {publisher} {Springer International Publishing},\ \bibinfo {address} {Heidelberg},\ \bibinfo {year} {2013})\ pp.\ \bibinfo {pages} {3--17}\BibitemShut {NoStop}%
\bibitem [{\citenamefont {Kivlichan}\ \emph {et~al.}(2018)\citenamefont {Kivlichan}, \citenamefont {McClean}, \citenamefont {Wiebe}, \citenamefont {Gidney}, \citenamefont {Aspuru-Guzik}, \citenamefont {Chan},\ and\ \citenamefont {Babbush}}]{Kivlichan2018}%
  \BibitemOpen
  \bibfield  {author} {\bibinfo {author} {\bibfnamefont {I.~D.}\ \bibnamefont {Kivlichan}}, \bibinfo {author} {\bibfnamefont {J.}~\bibnamefont {McClean}}, \bibinfo {author} {\bibfnamefont {N.}~\bibnamefont {Wiebe}}, \bibinfo {author} {\bibfnamefont {C.}~\bibnamefont {Gidney}}, \bibinfo {author} {\bibfnamefont {A.}~\bibnamefont {Aspuru-Guzik}}, \bibinfo {author} {\bibfnamefont {G.~K.-L.}\ \bibnamefont {Chan}},\ and\ \bibinfo {author} {\bibfnamefont {R.}~\bibnamefont {Babbush}},\ }\bibfield  {title} {\bibinfo {title} {{Quantum Simulation of Electronic Structure with Linear Depth and Connectivity}},\ }\href {https://doi.org/10.1103/PhysRevLett.120.110501} {\bibfield  {journal} {\bibinfo  {journal} {Physical Review Letters}\ }\textbf {\bibinfo {volume} {120}},\ \bibinfo {pages} {110501} (\bibinfo {year} {2018})}\BibitemShut {NoStop}%
\bibitem [{\citenamefont {O'Gorman}\ \emph {et~al.}(2019)\citenamefont {O'Gorman}, \citenamefont {Huggins}, \citenamefont {Rieffel},\ and\ \citenamefont {Whaley}}]{OGorman2019}%
  \BibitemOpen
  \bibfield  {author} {\bibinfo {author} {\bibfnamefont {B.}~\bibnamefont {O'Gorman}}, \bibinfo {author} {\bibfnamefont {W.~J.}\ \bibnamefont {Huggins}}, \bibinfo {author} {\bibfnamefont {E.~G.}\ \bibnamefont {Rieffel}},\ and\ \bibinfo {author} {\bibfnamefont {K.~B.}\ \bibnamefont {Whaley}},\ }\href {https://doi.org/10.48550/arXiv.1905.05118} {\bibinfo {title} {{Generalized swap networks for near-term quantum computing}}} (\bibinfo {year} {2019}),\ \bibinfo {note} {arXiv preprint arXiv:1905.05118}\BibitemShut {NoStop}%
\bibitem [{\citenamefont {Hagge}(2020)}]{Hagge2020}%
  \BibitemOpen
  \bibfield  {author} {\bibinfo {author} {\bibfnamefont {T.}~\bibnamefont {Hagge}},\ }\href {https://doi.org/10.48550/arXiv.2001.08324} {\bibinfo {title} {{Optimal fermionic swap networks for Hubbard models}}} (\bibinfo {year} {2020}),\ \bibinfo {note} {arXiv preprint arXiv:2001.08324}\BibitemShut {NoStop}%
\bibitem [{\citenamefont {Hashim}\ \emph {et~al.}(2022)\citenamefont {Hashim}, \citenamefont {Rines}, \citenamefont {Omole}, \citenamefont {Naik}, \citenamefont {Kreikebaum}, \citenamefont {Santiago}, \citenamefont {Chong}, \citenamefont {Siddiqi},\ and\ \citenamefont {Gokhale}}]{Hashim2022}%
  \BibitemOpen
  \bibfield  {author} {\bibinfo {author} {\bibfnamefont {A.}~\bibnamefont {Hashim}}, \bibinfo {author} {\bibfnamefont {R.}~\bibnamefont {Rines}}, \bibinfo {author} {\bibfnamefont {V.}~\bibnamefont {Omole}}, \bibinfo {author} {\bibfnamefont {R.~K.}\ \bibnamefont {Naik}}, \bibinfo {author} {\bibfnamefont {J.~M.}\ \bibnamefont {Kreikebaum}}, \bibinfo {author} {\bibfnamefont {D.~I.}\ \bibnamefont {Santiago}}, \bibinfo {author} {\bibfnamefont {F.~T.}\ \bibnamefont {Chong}}, \bibinfo {author} {\bibfnamefont {I.}~\bibnamefont {Siddiqi}},\ and\ \bibinfo {author} {\bibfnamefont {P.}~\bibnamefont {Gokhale}},\ }\bibfield  {title} {\bibinfo {title} {{Optimized SWAP networks with equivalent circuit averaging for QAOA}},\ }\href {https://doi.org/10.1103/PhysRevResearch.4.033028} {\bibfield  {journal} {\bibinfo  {journal} {Physical Review Research}\ }\textbf {\bibinfo {volume} {4}},\ \bibinfo {pages} {033028} (\bibinfo {year} {2022})}\BibitemShut {NoStop}%
\bibitem [{\citenamefont {Hen}\ and\ \citenamefont {Spedalieri}(2016)}]{Hen2016}%
  \BibitemOpen
  \bibfield  {author} {\bibinfo {author} {\bibfnamefont {I.}~\bibnamefont {Hen}}\ and\ \bibinfo {author} {\bibfnamefont {F.~M.}\ \bibnamefont {Spedalieri}},\ }\bibfield  {title} {\bibinfo {title} {{Quantum Annealing for Constrained Optimization}},\ }\href {https://doi.org/10.1103/PhysRevApplied.5.034007} {\bibfield  {journal} {\bibinfo  {journal} {Physical Review Applied}\ }\textbf {\bibinfo {volume} {5}},\ \bibinfo {pages} {034007} (\bibinfo {year} {2016})}\BibitemShut {NoStop}%
\bibitem [{\citenamefont {Hen}\ and\ \citenamefont {Sarandy}(2016)}]{Hen2016a}%
  \BibitemOpen
  \bibfield  {author} {\bibinfo {author} {\bibfnamefont {I.}~\bibnamefont {Hen}}\ and\ \bibinfo {author} {\bibfnamefont {M.~S.}\ \bibnamefont {Sarandy}},\ }\bibfield  {title} {\bibinfo {title} {{Driver Hamiltonians for constrained optimization in quantum annealing}},\ }\href {https://doi.org/10.1103/PhysRevA.93.062312} {\bibfield  {journal} {\bibinfo  {journal} {Physical Review A}\ }\textbf {\bibinfo {volume} {93}},\ \bibinfo {pages} {062312} (\bibinfo {year} {2016})}\BibitemShut {NoStop}%
\bibitem [{\citenamefont {Hadfield}\ \emph {et~al.}(2019)\citenamefont {Hadfield}, \citenamefont {Wang}, \citenamefont {O'Gorman}, \citenamefont {Rieffel}, \citenamefont {Venturelli},\ and\ \citenamefont {Biswas}}]{Hadfield2019}%
  \BibitemOpen
  \bibfield  {author} {\bibinfo {author} {\bibfnamefont {S.}~\bibnamefont {Hadfield}}, \bibinfo {author} {\bibfnamefont {Z.}~\bibnamefont {Wang}}, \bibinfo {author} {\bibfnamefont {B.}~\bibnamefont {O'Gorman}}, \bibinfo {author} {\bibfnamefont {E.}~\bibnamefont {Rieffel}}, \bibinfo {author} {\bibfnamefont {D.}~\bibnamefont {Venturelli}},\ and\ \bibinfo {author} {\bibfnamefont {R.}~\bibnamefont {Biswas}},\ }\bibfield  {title} {\bibinfo {title} {{From the Quantum Approximate Optimization Algorithm to a Quantum Alternating Operator Ansatz}},\ }\href {https://doi.org/10.3390/a12020034} {\bibfield  {journal} {\bibinfo  {journal} {Algorithms}\ }\textbf {\bibinfo {volume} {12}},\ \bibinfo {pages} {34} (\bibinfo {year} {2019})}\BibitemShut {NoStop}%
\bibitem [{\citenamefont {Drieb-Sch{\"{o}}n}\ \emph {et~al.}(2023)\citenamefont {Drieb-Sch{\"{o}}n}, \citenamefont {Ender}, \citenamefont {Javanmard},\ and\ \citenamefont {Lechner}}]{Drieb-Schon2023}%
  \BibitemOpen
  \bibfield  {author} {\bibinfo {author} {\bibfnamefont {M.}~\bibnamefont {Drieb-Sch{\"{o}}n}}, \bibinfo {author} {\bibfnamefont {K.}~\bibnamefont {Ender}}, \bibinfo {author} {\bibfnamefont {Y.}~\bibnamefont {Javanmard}},\ and\ \bibinfo {author} {\bibfnamefont {W.}~\bibnamefont {Lechner}},\ }\bibfield  {title} {\bibinfo {title} {{Parity Quantum Optimization: Encoding Constraints}},\ }\href {https://doi.org/10.22331/q-2023-03-17-951} {\bibfield  {journal} {\bibinfo  {journal} {Quantum}\ }\textbf {\bibinfo {volume} {7}},\ \bibinfo {pages} {951} (\bibinfo {year} {2023})}\BibitemShut {NoStop}%
\bibitem [{\citenamefont {Vyskocil}\ and\ \citenamefont {Djidjev}(2019)}]{Vyskocil2019}%
  \BibitemOpen
  \bibfield  {author} {\bibinfo {author} {\bibfnamefont {T.}~\bibnamefont {Vyskocil}}\ and\ \bibinfo {author} {\bibfnamefont {H.}~\bibnamefont {Djidjev}},\ }\bibfield  {title} {\bibinfo {title} {{Embedding equality constraints of optimization problems into a quantum annealer}},\ }\href {https://doi.org/10.3390/A12040077} {\bibfield  {journal} {\bibinfo  {journal} {Algorithms}\ }\textbf {\bibinfo {volume} {12}},\ \bibinfo {pages} {1} (\bibinfo {year} {2019})}\BibitemShut {NoStop}%
\bibitem [{\citenamefont {Vysko{\v{c}}il}\ \emph {et~al.}(2019)\citenamefont {Vysko{\v{c}}il}, \citenamefont {Pakin},\ and\ \citenamefont {Djidjev}}]{Vyskocil2019a}%
  \BibitemOpen
  \bibfield  {author} {\bibinfo {author} {\bibfnamefont {T.}~\bibnamefont {Vysko{\v{c}}il}}, \bibinfo {author} {\bibfnamefont {S.}~\bibnamefont {Pakin}},\ and\ \bibinfo {author} {\bibfnamefont {H.~N.}\ \bibnamefont {Djidjev}},\ }\bibfield  {title} {\bibinfo {title} {{Embedding Inequality Constraints for Quantum Annealing Optimization}},\ }in\ \href {https://doi.org/10.1007/978-3-030-14082-3_2} {\emph {\bibinfo {booktitle} {Quantum Technology and Optimization Problems}}},\ \bibinfo {series} {Lecture Notes in Computer Science}, Vol.\ \bibinfo {volume} {11413},\ \bibinfo {editor} {edited by\ \bibinfo {editor} {\bibfnamefont {S.}~\bibnamefont {Feld}}\ and\ \bibinfo {editor} {\bibfnamefont {C.}~\bibnamefont {Linnhoff-Popien}}}\ (\bibinfo  {publisher} {Springer International Publishing},\ \bibinfo {address} {Cham},\ \bibinfo {year} {2019})\ pp.\ \bibinfo {pages} {11--22}\BibitemShut {NoStop}%
\bibitem [{\citenamefont {Djidjev}(2020)}]{Djidjev2020}%
  \BibitemOpen
  \bibfield  {author} {\bibinfo {author} {\bibfnamefont {H.}~\bibnamefont {Djidjev}},\ }\bibfield  {title} {\bibinfo {title} {{Automaton-based methodology for implementing optimization constraints for quantum annealing}},\ }in\ \href {https://doi.org/10.1145/3387902.3392619} {\emph {\bibinfo {booktitle} {Proceedings of the 17th ACM International Conference on Computing Frontiers}}}\ (\bibinfo  {publisher} {ACM},\ \bibinfo {address} {New York, NY, USA},\ \bibinfo {year} {2020})\ pp.\ \bibinfo {pages} {118--125}\BibitemShut {NoStop}%
\bibitem [{\citenamefont {Fletcher}(1983)}]{Fletcher1983}%
  \BibitemOpen
  \bibfield  {author} {\bibinfo {author} {\bibfnamefont {R.}~\bibnamefont {Fletcher}},\ }\bibinfo {title} {{Penalty Functions}},\ in\ \href {https://doi.org/10.1007/978-3-642-68874-4_5} {\emph {\bibinfo {booktitle} {Mathematical Programming The State of the Art}}},\ \bibinfo {editor} {edited by\ \bibinfo {editor} {\bibfnamefont {A.}~\bibnamefont {Bachem}}, \bibinfo {editor} {\bibfnamefont {B.}~\bibnamefont {Korte}},\ and\ \bibinfo {editor} {\bibfnamefont {M.}~\bibnamefont {Gr{\"{o}}tschel}}}\ (\bibinfo  {publisher} {Springer Berlin Heidelberg},\ \bibinfo {address} {Berlin, Heidelberg},\ \bibinfo {year} {1983})\ pp.\ \bibinfo {pages} {87--114}\BibitemShut {NoStop}%
\bibitem [{\citenamefont {de~la Grand'rive}\ and\ \citenamefont {Hullo}(2019)}]{DelaGrandrive2019}%
  \BibitemOpen
  \bibfield  {author} {\bibinfo {author} {\bibfnamefont {P.~D.}\ \bibnamefont {de~la Grand'rive}}\ and\ \bibinfo {author} {\bibfnamefont {J.-F.}\ \bibnamefont {Hullo}},\ }\href {https://doi.org/10.48550/arXiv.1908.02210} {\bibinfo {title} {{Knapsack Problem variants of QAOA for battery revenue optimisation}}} (\bibinfo {year} {2019}),\ \bibinfo {note} {arXiv preprint arXiv:1908.02210}\BibitemShut {NoStop}%
\bibitem [{\citenamefont {Yu}\ and\ \citenamefont {Nabil}(2021)}]{Yu2021}%
  \BibitemOpen
  \bibfield  {author} {\bibinfo {author} {\bibfnamefont {S.}~\bibnamefont {Yu}}\ and\ \bibinfo {author} {\bibfnamefont {T.}~\bibnamefont {Nabil}},\ }\bibfield  {title} {\bibinfo {title} {{Applying the Hubbard-Stratonovich Transformation to Solve Scheduling Problems Under Inequality Constraints With Quantum Annealing}},\ }\href {https://doi.org/10.3389/fphy.2021.730685} {\bibfield  {journal} {\bibinfo  {journal} {Frontiers in Physics}\ }\textbf {\bibinfo {volume} {9}},\ \bibinfo {pages} {1} (\bibinfo {year} {2021})}\BibitemShut {NoStop}%
\bibitem [{\citenamefont {Stratonovich}(1957)}]{stratonovich1957method}%
  \BibitemOpen
  \bibfield  {author} {\bibinfo {author} {\bibfnamefont {R.~L.}\ \bibnamefont {Stratonovich}},\ }\bibfield  {title} {\bibinfo {title} {{On a method of calculating quantum distribution functions}},\ }in\ \href@noop {} {\emph {\bibinfo {booktitle} {Soviet Physics Doklady}}},\ Vol.~\bibinfo {volume} {2}\ (\bibinfo {year} {1957})\ p.\ \bibinfo {pages} {416}\BibitemShut {NoStop}%
\bibitem [{\citenamefont {Hubbard}(1959)}]{Hubbard1959}%
  \BibitemOpen
  \bibfield  {author} {\bibinfo {author} {\bibfnamefont {J.}~\bibnamefont {Hubbard}},\ }\bibfield  {title} {\bibinfo {title} {{Calculation of Partition Functions}},\ }\href {https://doi.org/10.1103/PhysRevLett.3.77} {\bibfield  {journal} {\bibinfo  {journal} {Physical Review Letters}\ }\textbf {\bibinfo {volume} {3}},\ \bibinfo {pages} {77} (\bibinfo {year} {1959})}\BibitemShut {NoStop}%
\bibitem [{\citenamefont {Kuramata}\ \emph {et~al.}(2021)\citenamefont {Kuramata}, \citenamefont {Katsuki},\ and\ \citenamefont {Nakata}}]{Kuramata2021}%
  \BibitemOpen
  \bibfield  {author} {\bibinfo {author} {\bibfnamefont {M.}~\bibnamefont {Kuramata}}, \bibinfo {author} {\bibfnamefont {R.}~\bibnamefont {Katsuki}},\ and\ \bibinfo {author} {\bibfnamefont {K.}~\bibnamefont {Nakata}},\ }\bibfield  {title} {\bibinfo {title} {{Larger Sparse Quadratic Assignment Problem Optimization Using Quantum Annealing and a Bit-Flip Heuristic Algorithm}},\ }in\ \href {https://doi.org/10.1109/ICIEA52957.2021.9436749} {\emph {\bibinfo {booktitle} {2021 IEEE 8th International Conference on Industrial Engineering and Applications (ICIEA)}}}\ (\bibinfo  {publisher} {IEEE},\ \bibinfo {year} {2021})\ pp.\ \bibinfo {pages} {556--565}\BibitemShut {NoStop}%
\bibitem [{\citenamefont {van Rossum}\ and\ \citenamefont {{Drake Jr}}(1995)}]{VanRossum1995}%
  \BibitemOpen
  \bibfield  {author} {\bibinfo {author} {\bibfnamefont {G.}~\bibnamefont {van Rossum}}\ and\ \bibinfo {author} {\bibfnamefont {F.~L.}\ \bibnamefont {{Drake Jr}}},\ }\href@noop {} {\emph {\bibinfo {title} {{Python tutorial}}}}\ (\bibinfo  {publisher} {Centrum voor Wiskunde en Informatica},\ \bibinfo {year} {1995})\BibitemShut {NoStop}%
\bibitem [{\citenamefont {Harris}\ \emph {et~al.}(2020)\citenamefont {Harris}, \citenamefont {Millman}, \citenamefont {van~der Walt}, \citenamefont {Gommers}, \citenamefont {Virtanen}, \citenamefont {Cournapeau}, \citenamefont {Wieser}, \citenamefont {Taylor}, \citenamefont {Berg}, \citenamefont {Smith} \emph {et~al.}}]{harris2020array}%
  \BibitemOpen
  \bibfield  {author} {\bibinfo {author} {\bibfnamefont {C.~R.}\ \bibnamefont {Harris}}, \bibinfo {author} {\bibfnamefont {K.~J.}\ \bibnamefont {Millman}}, \bibinfo {author} {\bibfnamefont {S.~J.}\ \bibnamefont {van~der Walt}}, \bibinfo {author} {\bibfnamefont {R.}~\bibnamefont {Gommers}}, \bibinfo {author} {\bibfnamefont {P.}~\bibnamefont {Virtanen}}, \bibinfo {author} {\bibfnamefont {D.}~\bibnamefont {Cournapeau}}, \bibinfo {author} {\bibfnamefont {E.}~\bibnamefont {Wieser}}, \bibinfo {author} {\bibfnamefont {J.}~\bibnamefont {Taylor}}, \bibinfo {author} {\bibfnamefont {S.}~\bibnamefont {Berg}}, \bibinfo {author} {\bibfnamefont {N.~J.}\ \bibnamefont {Smith}}, \emph {et~al.},\ }\bibfield  {title} {\bibinfo {title} {{Array programming with NumPy}},\ }\href {https://doi.org/10.1038/s41586-020-2649-2} {\bibfield  {journal} {\bibinfo  {journal} {Nature}\ }\textbf {\bibinfo {volume} {585}},\ \bibinfo {pages} {357} (\bibinfo {year} {2020})}\BibitemShut {NoStop}%
\bibitem [{\citenamefont {Virtanen}\ \emph {et~al.}(2020)\citenamefont {Virtanen}, \citenamefont {Gommers}, \citenamefont {Oliphant}, \citenamefont {Haberland}, \citenamefont {Reddy}, \citenamefont {Cournapeau}, \citenamefont {Burovski}, \citenamefont {Peterson}, \citenamefont {Weckesser}, \citenamefont {Bright} \emph {et~al.}}]{2020SciPy-NMeth}%
  \BibitemOpen
  \bibfield  {author} {\bibinfo {author} {\bibfnamefont {P.}~\bibnamefont {Virtanen}}, \bibinfo {author} {\bibfnamefont {R.}~\bibnamefont {Gommers}}, \bibinfo {author} {\bibfnamefont {T.~E.}\ \bibnamefont {Oliphant}}, \bibinfo {author} {\bibfnamefont {M.}~\bibnamefont {Haberland}}, \bibinfo {author} {\bibfnamefont {T.}~\bibnamefont {Reddy}}, \bibinfo {author} {\bibfnamefont {D.}~\bibnamefont {Cournapeau}}, \bibinfo {author} {\bibfnamefont {E.}~\bibnamefont {Burovski}}, \bibinfo {author} {\bibfnamefont {P.}~\bibnamefont {Peterson}}, \bibinfo {author} {\bibfnamefont {W.}~\bibnamefont {Weckesser}}, \bibinfo {author} {\bibfnamefont {J.}~\bibnamefont {Bright}}, \emph {et~al.},\ }\bibfield  {title} {\bibinfo {title} {{SciPy 1.0: fundamental algorithms for scientific computing in Python}},\ }\href {https://doi.org/10.1038/s41592-019-0686-2} {\bibfield  {journal} {\bibinfo  {journal} {Nature Methods}\ }\textbf {\bibinfo {volume} {17}},\ \bibinfo {pages} {261} (\bibinfo {year} {2020})}\BibitemShut {NoStop}%
\bibitem [{\citenamefont {Zaman}\ \emph {et~al.}(2022)\citenamefont {Zaman}, \citenamefont {Tanahashi},\ and\ \citenamefont {Tanaka}}]{Zaman2022}%
  \BibitemOpen
  \bibfield  {author} {\bibinfo {author} {\bibfnamefont {M.}~\bibnamefont {Zaman}}, \bibinfo {author} {\bibfnamefont {K.}~\bibnamefont {Tanahashi}},\ and\ \bibinfo {author} {\bibfnamefont {S.}~\bibnamefont {Tanaka}},\ }\bibfield  {title} {\bibinfo {title} {{PyQUBO: Python Library for Mapping Combinatorial Optimization Problems to QUBO Form}},\ }\href {https://doi.org/10.1109/TC.2021.3063618} {\bibfield  {journal} {\bibinfo  {journal} {IEEE Transactions on Computers}\ }\textbf {\bibinfo {volume} {71}},\ \bibinfo {pages} {838} (\bibinfo {year} {2022})}\BibitemShut {NoStop}%
\bibitem [{\citenamefont {Hunter}(2007)}]{hunter2007matplotlib}%
  \BibitemOpen
  \bibfield  {author} {\bibinfo {author} {\bibfnamefont {J.~D.}\ \bibnamefont {Hunter}},\ }\bibfield  {title} {\bibinfo {title} {{Matplotlib: A 2D Graphics Environment}},\ }\href {https://doi.org/10.1109/MCSE.2007.55} {\bibfield  {journal} {\bibinfo  {journal} {Computing in Science \& Engineering}\ }\textbf {\bibinfo {volume} {9}},\ \bibinfo {pages} {90} (\bibinfo {year} {2007})}\BibitemShut {NoStop}%
\bibitem [{\citenamefont {{Gurobi Optimization, LLC}}(2023)}]{gurobi}%
  \BibitemOpen
  \bibfield  {author} {\bibinfo {author} {\bibnamefont {{Gurobi Optimization, LLC}}},\ }\href {https://www.gurobi.com} {\bibinfo {title} {{Gurobi Optimizer Reference Manual}}} (\bibinfo {year} {2023})\BibitemShut {NoStop}%
\bibitem [{\citenamefont {Callison}\ \emph {et~al.}(2021)\citenamefont {Callison}, \citenamefont {Festenstein}, \citenamefont {Chen}, \citenamefont {Nita}, \citenamefont {Kendon},\ and\ \citenamefont {Chancellor}}]{Festenstein}%
  \BibitemOpen
  \bibfield  {author} {\bibinfo {author} {\bibfnamefont {A.}~\bibnamefont {Callison}}, \bibinfo {author} {\bibfnamefont {M.}~\bibnamefont {Festenstein}}, \bibinfo {author} {\bibfnamefont {J.}~\bibnamefont {Chen}}, \bibinfo {author} {\bibfnamefont {L.}~\bibnamefont {Nita}}, \bibinfo {author} {\bibfnamefont {V.}~\bibnamefont {Kendon}},\ and\ \bibinfo {author} {\bibfnamefont {N.}~\bibnamefont {Chancellor}},\ }\bibfield  {title} {\bibinfo {title} {{Energetic Perspective on Rapid Quenches in Quantum Annealing}},\ }\href {https://doi.org/10.1103/PRXQuantum.2.010338} {\bibfield  {journal} {\bibinfo  {journal} {PRX Quantum}\ }\textbf {\bibinfo {volume} {2}},\ \bibinfo {pages} {010338} (\bibinfo {year} {2021})}\BibitemShut {NoStop}%
\bibitem [{\citenamefont {{Qiskit contributors}}(2023)}]{Qiskit}%
  \BibitemOpen
  \bibfield  {author} {\bibinfo {author} {\bibnamefont {{Qiskit contributors}}},\ }\href {https://doi.org/10.5281/zenodo.2573505} {\bibinfo {title} {Qiskit: An open-source framework for quantum computing}} (\bibinfo {year} {2023})\BibitemShut {NoStop}%
\bibitem [{\citenamefont {Powell}(1994)}]{Powell1994}%
  \BibitemOpen
  \bibfield  {author} {\bibinfo {author} {\bibfnamefont {M.~J.~D.}\ \bibnamefont {Powell}},\ }\bibinfo {title} {{A Direct Search Optimization Method That Models the Objective and Constraint Functions by Linear Interpolation}},\ in\ \href {https://doi.org/10.1007/978-94-015-8330-5_4} {\emph {\bibinfo {booktitle} {Advances in Optimization and Numerical Analysis}}},\ \bibinfo {editor} {edited by\ \bibinfo {editor} {\bibfnamefont {S.}~\bibnamefont {Gomez}}\ and\ \bibinfo {editor} {\bibfnamefont {J.-P.}\ \bibnamefont {Hennart}}}\ (\bibinfo  {publisher} {Springer Netherlands},\ \bibinfo {address} {Dordrecht},\ \bibinfo {year} {1994})\ pp.\ \bibinfo {pages} {51--67}\BibitemShut {NoStop}%
\bibitem [{\citenamefont {Callison}\ and\ \citenamefont {Chancellor}(2022)}]{Callison2022}%
  \BibitemOpen
  \bibfield  {author} {\bibinfo {author} {\bibfnamefont {A.}~\bibnamefont {Callison}}\ and\ \bibinfo {author} {\bibfnamefont {N.}~\bibnamefont {Chancellor}},\ }\bibfield  {title} {\bibinfo {title} {Hybrid quantum-classical algorithms in the noisy intermediate-scale quantum era and beyond},\ }\href {https://doi.org/10.1103/PhysRevA.106.010101} {\bibfield  {journal} {\bibinfo  {journal} {Phys. Rev. A}\ }\textbf {\bibinfo {volume} {106}},\ \bibinfo {pages} {010101} (\bibinfo {year} {2022})}\BibitemShut {NoStop}%
\bibitem [{\citenamefont {Mirkarimi}\ \emph {et~al.}(2024{\natexlab{b}})\citenamefont {Mirkarimi}, \citenamefont {Shukla}, \citenamefont {Hoyle}, \citenamefont {Williams},\ and\ \citenamefont {Chancellor}}]{Mirkarimi2024data}%
  \BibitemOpen
  \bibfield  {author} {\bibinfo {author} {\bibfnamefont {P.}~\bibnamefont {Mirkarimi}}, \bibinfo {author} {\bibfnamefont {I.}~\bibnamefont {Shukla}}, \bibinfo {author} {\bibfnamefont {D.~C.}\ \bibnamefont {Hoyle}}, \bibinfo {author} {\bibfnamefont {R.}~\bibnamefont {Williams}},\ and\ \bibinfo {author} {\bibfnamefont {N.}~\bibnamefont {Chancellor}},\ }\href {https://doi.org/10.15128/r2fq977t82m} {\bibinfo {title} {{Quantum optimization with linear Ising penalty functions for customer data science [dataset]}}} (\bibinfo {year} {2024}{\natexlab{b}}),\ \bibinfo {note} {data archive at Durham University, UK}\BibitemShut {NoStop}%
\end{thebibliography}
\end{document}